# Multimodal phase velocity-frequency dispersion images using different MASW transformation techniques

Jyant Kumar, Tarun Naskar[*],

*Indian Institute of Science, Bengaluru-560012, India*

*Indian Institute of Technology, Madras-600036, India*

**Abstract**

Three different transformation techniques, namely, (i) $\omega$–$c$, (ii) $\omega$–$\kappa$, and (iii) $\tau$–$p$, has been employed for generating multimodal dispersion images on the basis of multi-channel analysis of surface waves (MASW) data recorded in distance-time domain; here $\omega$ = circular frequency, $c$ = phase velocity, $\tau$ = time intercept, $p$ = phase slowness ($1/c$) and $\kappa$ = wavenumber. All the three methods have been first clearly described. The results from these three different transforms have been examined by using synthetic as well as field data obtained from field tests using 48 geophones. The effect of sensor spread length ($X$) and geophone numbers ($M$) on multimodal dispersion images were examined. The solutions from these three transforms were found to match generally well with each other. The $\omega$–$c$ transform has been noted to provide the most clarity since it does not require either high sampling rate as normally needed for the $\tau$-$p$ method or inclusion of the zero padding of the data in a distance domain for the $\omega$-$k$ approach. The paper will be useful since it not only describes the methods, but it brings out simultaneously their merits in implementation and for generating the dispersion images.

[*]Corresponding author. Tel.: +91 44 2257 4322;.

Email addresses: jkumar@iisc.ac.in (J. Kumar); tarunnaskar@iitm.ac.in(T. Naskar)

*Keywords*: Spectral analysis; MASW; Dispersion; Fourier transform; Wave propagation

## 1. Introduction

Surface wave testing methods, namely, (i) spectral analysis of surface waves (SASW), and (ii) multichannel analysis of surface waves (MASW) are two non-intrusive and non-destructive techniques which are often employed to assess the shear wave velocity profile of ground and pavement sites (Nazarian and Desai, 1993; Park et al., 1998, 1999; Ivanov et. Al., 2011; Lin, 2014 Naskar, 2017). These techniques are quite cost-effective and can be implemented very easily. Surface wave testing techniques constitute mainly three different aspects (i) acquisition of field vibration data usually by means of geophones, (ii) obtaining field dispersion plots, and (iii) performing an inversion analysis by matching the obtained field dispersion curve with that computed from the forward analysis so that an accurate prediction of the layered media can be ascertained (Kausel and Roësset, 1981; Gucunski and Woods, 1992; Tokimatsu et al. 1992; Zomorodian et al., 2006; Kumar and Naskar, 2015, 2017a; Naskar and Kumar, 2017). In the SASW tests, usually two to six sensors are employed for the procurement of the field data, and only the predominant/superposed mode of propagation can be noted in the dispersion plot (Tokimatsu et al., 1992). However, it is well understood that an interference of different modes can significantly vary the dispersion plots, and therefore, a prediction simply based on the employment of the predominant/superposed mode may lead to an erroneous result especially for an irregular dispersive ground profile (Kausel and Roësset, 1981; Tokimatsu et al., 1992; Naskar and Kumar, 2017). On the other hand, depending upon acquisition geometry and velocity model, the MASW method can identify different modes to generate the experimental dispersion graph (Park et al., 1999).

Three different transformation methods, namely, (i) $\omega$–$c$ (Park et al., 1998), (ii) $\omega$–$\kappa$ (Yilmaz, 1987; Gabriels et al., 1987) and (iii) $\tau$–$p$ (McMechan and Yedlin, 1981) are used for generating the multimodal dispersion plots; here $\omega$ = circular frequency, $c$ = phase velocity, $\kappa$ = wavenumber, $\tau$ = time intercept, $p$ = phase slowness ($1/c$). In the present paper, dispersion images has been generated from these three different transformation methods with the usage of synthetic as well as actual experimental data which was generated by employing 48 geophones for two different ground sites. In the initial section, three transformation methods has been clearly described. The effect of the (i) sensor spread length (*X*), and (ii) the number of channels (*M*) employed for the recording of data in the distance domain were analyzed. The objective of the study is to assess the quality of the phase velocity-frequency dispersion images which can be determined with the usage of each of the three transformation methods.

## 2. Different transformation methods

### 2.1. Circular frequency ($\omega$) - phase velocity (c) transform

Let $u(x,t)$ be the collection of data in distance (*x*) - time (*t*) domain. Then transforming the equation to $\omega$-*c* domain, the following equation is formed (Park et al., 1998):

$$W(\omega, c) = \int_{-\infty}^{+\infty} e^{-i\left(\frac{\omega}{c}\right)x} \frac{U(x, \omega)}{|U(x, \omega)|} dx \qquad (1)$$

where $\kappa = \frac{\omega}{c} = \frac{2\pi}{\lambda}$; $\lambda$ = wavelength; and *c* = the phase velocity corresponding to the circular frequency $\omega$. This expression in a discrete form for distance interval $\Delta x$ with *M* records in distance domain becomes as:

$$W(\omega, c) = \sum_{j=1}^{M} e^{-i\left(\frac{\omega}{c}\right)x_j} \frac{U(x_j, \omega)}{|U(x_j, \omega)|} \Delta x \qquad (2)$$

By using Eq. (2), one can generate phase velocity- frequency image with the $\omega$ representing the *x*-axis and *c* representing *y*-axis. The value of the function $W(\omega, c)$ for a particular circular frequency ($\omega$) becomes maximum only for a specific value of *c*. By determining all the maxima for the function $W(\omega, c)$, a relationship between $\omega$ and *c* can, therefore, be determined. On this basis, one can generate the dispersion plots associated with the multiple modes.

### 2.2. Circular frequency ($\omega$) - wavenumber ($\kappa$) transform

In this method, a two-dimensional Fourier transform is performed to transform the recorded data from the *x-t* domain to the $\omega$-$\kappa$ domain (Yilmaz, 1987; Gabriels et al., 1987)

$$U(\omega, \kappa) = \int_{-\infty}^{\infty} \int_{-\infty}^{\infty} u(x, t) e^{-i(\omega t - \kappa x)} dx \, dt \qquad (3)$$

The discrete form of this expression becomes as:

$$U(\omega, \kappa) = \sum_{l=1}^{N} \sum_{j=1}^{M} u(x_j, t_l) e^{-i(\omega t_l - \kappa x_j)} \Delta x \Delta t \tag{4}$$

Again, one can obtain the dispersion plot by finding the value of $\kappa$ for a particular $\omega$ for which the magnitude of $U$ becomes maximum.

In this transformation technique, the number of points along the distance (wavenumber) domain needs to be increased using zero padding ($\Lambda$). It implies that some virtual channels need to be inserted either before the first sensor or beyond the last sensor. Each virtual channel is assumed to generate null (zero) data record by keeping its total time duration and the data sampling rate to be the same as that of the actual sensor. In the present analysis, 25X - 50X zero padding was adopted which implies that the total numbers of virtual channels were kept twenty-five to fifty times respectively as that of the actual number of chosen sensors. For example, let assume in a particular test, total $n$ no of sensors were employed to record a signal. Then, 25X of zero padding to that signal implies that now there will be $n \times 25 = 25n$ effectual sensors. 1X zero padding would mean no virtual sensors were added to the number of actual sensors. It should be mentioned that generally, no issue arises for increasing the resolution towards the frequency which can be simply achieved by choosing a high sampling rate and increased recording duration. However, in most of the cases, the numbers of sensors are limited, and as a result, the zero padding of the data in spatial domain becomes essential to provide a better resolution in the wavenumber.

## 2.3. *Time intercept ($\tau$) – phase slowness (p) transform*

This transform allows the decomposition of the recorded signal in *x-t* domain into $\tau$ - $p$ domain, where $\tau$ refers to the time intercept and $p$ is the phase slowness, that is, $p = \frac{1}{c} = \frac{dt}{dx}$ (McMechan and Yedlin, 1981). The variables $\tau$ and $p$ are related with $x$ and $t$ by using the expression, $t = \tau + px$. The transform of the recorded data $u(x,t)$ in $\tau$-$p$ domain becomes as (McMechan and Yedlin, 1981):

$$f(\tau, p) = \int_{-\infty}^{\infty} u(x, \tau + px) dx \tag{5}$$

In a discrete form, this integral expression is given by

$$f(\tau, p) = \sum_{j=1}^{M} u(x_j, \tau + px_j) \Delta x \tag{6}$$

For a given value of $p$, $\Delta \tau = (\Delta t - p \Delta x)$. Therefore, when the function is plotted in $\tau$-$p$ domain, $\Delta \tau$ will become constant for a given $p$. After applying the Fourier transform over the variable $\tau$ for the function $f(\tau, p)$, the transformed function $F(\omega, p)$ is computed:

$$F(\omega, p) = \int_{-\infty}^{\infty} f(\tau, p) e^{-i\omega \tau} d\tau \tag{7}$$

In a discrete form, this expression becomes as:

$$F(\omega, p) = \sum_{j=1}^{N} f(\tau_j, p) e^{-i\omega \tau_j} \Delta \tau \tag{8}$$

The dispersion plot can be likewise generated by determining the value of $\omega$ for a given $p$ for which the magnitude of $F$ becomes maximum.

## 3. Data record

To assess the quality of dispersion plots obtained from the three different transforms, synthetic signals, as well as the actual experimental data records, were being utilized:

### 3.1. *Generation of synthetic data*

For generating the synthetic signal, first, a two-layer system followed by an elastic half-space with pre-assumed dynamic properties were selected (Fig. 1a). These parameters then used as inputs to generate theoretical dispersion graph (Fig. 1b) by root search method using LU decomposition (Naskar & Kumar, 2018) technique. The dispersion value obtained then utilized to generate time dependent synthetic signal for each sensor (Park et al. 2008). The synthetic signal comprising of P-SV components contains different modes for Rayleigh wave propagation. For validation purpose, all the modes were considered and their relative amplitudes were kept unity.

The synthetic data was generated for *M*=24, 48 and 96 channels by keeping the source to first sensor distance (*S*) 5 m and with a sensor spread length (*X*) of 46 m, 92 m and 184 m between first and last sensors. The method was described elaborately in Naskar and Kumar (2017).

### 3.2. *Experimental data*

For generating the actual field data, two sites were selected. The first site is located near Hebbal, Bengaluru and the second site was selected is an open field inside the campus of *Indian Institute of Science*, *Bengaluru*. A total number of *M* = 48 geophones with a natural frequency of 4.5 Hz, were employed by keeping the spacing between adjacent sensors ($\Delta x$) equal to 0.5 m. The source to first sensor distance (*S*) was kept equal to 2.5m. The data was collected with a sampling rate of 8 kHz and 48 kHz for Site-1 and Site-2 respectively. The total duration of the signal was kept equal to 5 sec in the ime domain. A circular steel base plate was placed on the surface of the ground and the vibrations were induced by hitting this plate by using a 20 lbs sledgehammer.

## 4. Analysis

### 4.1. Validation

For validation, dispersion images generated by all the three different methods were compared with the input model dispersion graph. Fig. 2 presents a comparison of the dispersion plots by using synthetic data from *M* =1000 channels; the sensors were spaced at an interval ($\Delta x$) of 2 m. The data sampling rate was kept equal to 1000 samples/sec except for $\tau$–$p$, method. Due to the $\tau$–$p$ method's susceptibility to the data sampling rate, it was kept equal to 25k samples/sec. similarly, for the $\omega$–$\kappa$ transform, $\Lambda$=50X data padding was employed to obtain better results. The requirement for these special treatments for these two methods are elaborated later in this paper. It can be seen that all the three methods tend to predict different modes of wave propagation more or less in a similar fashion. These three methods generated the dispersion plots for all the seven modes with more or less equal clarity. The $\omega$-$c$ transform produced the clearest and sharpest image among the three methods.

*4.2. Dispersion images using synthetic data*

The following sections present the results by three different transformation methods for synthetic as well as an actual field data record. Although all the seven modes were considered for validation purpose in the previous paragraph, in an actual field test, however, one can hardly observe more than 2 to 3 modes, thus considering all the higher modes deems to be unnecessarily complicating the dispersion plots. Also, in the field, the strength of all those modes will not remain constant, and they will vary with different frequencies. Therefore, apart from the validation, for all other purposes only first three modes were considered for generating synthetic signal and their amplitude ratio for different frequencies were calculated from input parameters (Naskar and Kumar, 2017) (Fig. 3); wherein the amplitude ratio defines the ratio of the displacement amplitude of the signal of a given mode to that with the predominant (fundamental) mode. Fig. 4-16 are associated with the usage of synthetic signals with the first three modes and corresponding relative amplitude ratios, on the other hand, Figs. 17-19 represents the data collected from actual field measurements.

Fig. 4-6 represents a comparison of the dispersion plots by three methods using synthetic data from 24, 48 and 96 channels. The sensors spread length were kept as $X=46$ m. For $\omega$–$c$ and $\omega$–$\kappa$ transform the data sampling rate was kept equal to 1 k/sec and for $\tau$–$p$ transform data sampling rate was kept equal to 100k samples/sec. For the $\omega$-$\kappa$ transform, 50X data padding was employed. It can be seen that all the three methods tend to predict different modes of wave propagation more or less in a similar fashion. All the methods generated the dispersion plots for the fundamental mode with better clarity. On the other hand, the dispersion curve associated with the higher modes becomes relatively weaker. The increment in the number of channels does not improve the dispersion image qualities. From the dispersion images generated by $\omega$-$\kappa$ transform, we can see an inclined (Strobbia, 2003) line along the frequency axis below which no information is available (Figs. 6). It is due to a phenomenon called the Nyquist sampling criterion which is explained in the appendix.

Fig. 7-9 represents the dispersion images with sensor spread length kept equal to $X$= 92 m. Compare to 46 m spread length (Fig 4-6), the dispersion images generated for the present scenario were found to be superior for all the three methods. All the methods produced the dispersion plots with more coherent higher modes. The role of number of channels ($M$) employed was found to be negligible and spread length ($X$) primarily dictates the image qualities. The $\omega$-$c$ transform produced marginally sharper and clearer dispersion images in a frequency band of 15-40 Hz as compared to the other two methods. On the other hand, no methods yield very clear dispersion images for the lower frequency range (0-15 Hz). The $\omega$-$c$ transform does not consider the variation in the amplitude for any particular mode over the different frequencies. Thus, when nonuniform amplitude was given as input (Fig 3), it overestimates the first mode's strength in the frequency range of 0-5Hz (Fig 4 and Fig 7). The $\omega$–$\kappa$ and $\tau$–$p$ methods were correct to predicts a very low amplitude first mode in the range of 0-15 Hz.

### 4.3. Difficulties associated with higher spread length (X)

So far it has been observed that the sensor spread length greatly improves the dispersion image quality for all the three methods and number of sensor played a trivial role in it. Fig. 10-12 represents dispersion images by three methods using spread length of 184 m with 24, 48 and 96 sensors. Table-1-2 represents the minimum frequency that can be roughly estimated with acceptable precision by all three methods using a combination of different spread length and number of sensors. It should be mentioned that, minimum frequency estimation for all the tables were done by manual judgment. Ability to determine phase velocity of lower frequencies with high precision helps determine deeper subsurface properties more accurately. From the Table. 1 it is clear that increasing sensor spread length helps estimating phase velocity of lower frequencies. On the other hand, from Table-2 it is clear that an increasing number of channels/sensors does not provide a better estimation of phase velocity of minimum frequencies. But increasing sensor spacing excessively for determining phase velocity of lower and lower frequencies may not be advisable as it can produced erroneous results. Fig 10a, 11a and 12a represents the difficulties associated when large sensor spacing was employed. When spread length was kept at 184 m using 24 geophones, the sensor spacing becomes 8 m which are too large. Thus, phase unwrapping problem (Kumar and Naskar, 2017b) kicks in and all the methods produced ghost modes i.e. the modes which do not exist in the input model curve (Fig 10a, 11a and 12a). The phase unwrapping problem depends on frequency and phase velocity of a particular modes thus difficult to predict in advance. On the other hand, for 48 and 96 channels, the sensor spacing got reduced thus it does not produce any ghost modes (Fig 10b,c, 11b,c and 12b,c). These types of problem are quite common in the different field of research and interpolation technique were often used to increase the resolution of the same. So, we investigated the efficacy of interpolation technique to increase the number of virtual sensors in the spatial domain. The smooth spline interpolation technique was used to increase the number of channels keeping the spread length fixed. Fig. 13a presents the synthetic offset data using 24 sensors, Fig. 13b represents offset data using 96 sensors and Fig. 13c represents offset data that produced for 24 sensors but interpolated to 96 sensors using smooth spline interpolation technique. For a casual observer, it is clear that Fig. 13b and Fig. 13c does not match at all. Figure 14 represents dispersion images produced from all these three offset data respectively. It is clear that a smooth spline interpolation technique not just only fails to eliminate ghost modes but produced inaccurate results (Fig 14c) which may lead to the incorrect prediction of soil properties.

### 4.4. Effect of sampling rate on the dispersion plots using the $\tau$-p method

Among the three methods, the sampling rate of the data in the temporal domain has a maximum effect on the dispersion plot generated by the $\tau$-p method. To examine this aspect, four different values of the sampling rates ($F_s$), namely, 5 kHz, 10 kHz, 25 kHz and 100 kHz were selected. Corresponding dispersion plots generated by the $\tau$-p method are illustrated in Fig. 15. It can be noted that the different modal curves tend to become sharper and clearer with an increase in the value of $F_s$. At around 25 KHz these improvements reached to its maximum effect and further increment in the $F_s$ only marginally improves the quality. In other words, if the data is recorded at a higher sampling rate, the $\tau$-$p$ transformation method will tend to reveal a better picture of the dispersion plots. It should be mentioned that we found no issues regarding smooth spline interpolation in case of temporal data and it can be used for increasing sampling rate.

### 4.5. Effect of zero data padding of the data using the $\omega$-$\kappa$ transform

As mentioned earlier, on account of a limited number of channels, it is often required that the zero padding of the data in a distance domain should be included. The data padding has maximum influence on the dispersion plots drawn with the usage of the $\omega$-$\kappa$ transform. The zero padding of the data can be easily included for the synthetic as well as actual experimental data. In Fig. 16, by the $\omega$-$\kappa$ transform, the effect of zero data padding has been illustrated with four different values, namely, $1X$, $5X$, $25X$ and $50X$ using 96 channels; the term $1X$ implies the record without zero data padding. It can be noted that the dispersion images tend to become clearer with an increase in the value of zero data padding. This finding is in agreement with the similar observation of Strobbia (2003). It should be mentioned that the $\omega$-$c$ and $\tau$-$p$ transforms are hardly affected by the inclusion of zero data padding. This aspect is, therefore, not included for the transforms by $\omega$-$c$ and $\tau$-$p$ techniques. It should be mentioned that because of the requirement of (i) zero data padding in $\omega$-$\kappa$ transform, and (ii) greater sampling rate for the $\tau$-$p$ method, these two methods generally require greater computational time as compared to the $\omega$-$c$ transform.

### 4.6. Dispersion plots using field data

For the actual test data obtained from the two chosen ground site, Figs. 17 and 18 illustrate dispersion images by the three different transforms. Fig 17 represents Site-1 which is located near Hebbal, Bengaluru India. This is an upcoming construction site. All the three methods predicted only one mode and $\omega$-$c$ transform produced sharpest of the three. An inversion analysis also performed for the Site -1 and results were presented in Fig. 17(d).

The Site-2 an open field inside the IISc campus Bengaluru, India. The site is located very close to an airfield thus highly irregular subsurface profile was expected. Fig. 18 represents the dispersion images for site-2. As anticipated, the dispersion images contain multiple higher modes representing irregular soil profile. Fig. 18(d) illustrates a combined comparison of the dispersion plots obtained by using all the three methods. Four different modes have been identified. For third and fourth modes, all the methods provide more or less the same unique dispersion relationship. For the first and second modes, between 75-100 Hz, all the methods compare quite favorably with each other. For the lower frequency range (<50 Hz), the $\omega$-$c$ transform provides the clearest dispersion relationship. On the other hand, for frequency <50 Hz, the $\omega$-$\kappa$ and $\tau$-$p$ transforms are unable to reveal clear information about the dispersion relationship. Fig 19 represents the inversion analysis data for Site-2. The highly irregular profile obtained from inverse analysis explains the reason behind dispersion image with multiple higher modes, thus bolstering the accuracy of the present analysis.

## 5. Conclusions

Three different transformation techniques, namely, (i) $\omega$-$c$, (ii) $\omega$-$\kappa$, and (iii) $\tau$-$p$ has been used to examine the dispersion relationship between the phase velocity and frequency. Both synthetic and the actual MASW data records obtained from the ground sites for different spread length and channel number were analyzed. For validation, dispersion images using 1000 channels were also generated. All the three methods compare favorably with each other. Effect of number of channels employed on dispersion images found to be trivial. The sensor spread length dictated the quality of dispersion images when sensor spacing was relatively small. The large sensor spacing can impart unwanted complicacy especially for higher frequencies and higher modes. The clarity of the dispersion relationship for the $\omega$-$\kappa$ method can be improved with the insertion of zero padding of the data in a distance (wavenumber) domain. For the $\tau$-$p$ transform, the dispersion relationship can be generated with more clarity by using a higher value of the data sampling rate.

In all the cases, the *ω-c* transform generally provides the sharpest phase velocity-frequency dispersion image associated with different modes. However, it overestimates the amplitude in case of a weaker predominant mode. Since *ω-c* transform does not require either zero padding of the data in a distance domain or very high sampling time interval, the method is computationally more economical to implement.

## 6. Acknowledgments

The authors wish to whole heartedly thank an anonymous reviewer for his/her valuable inputs. The financial support provided by (i) Department of Science and Technology, India (Grant No # SR/S3/MERC/072/2010) and (ii) Department of Atomic Energy, BRNS (Grant No # 2012/36/22-BRNS/1589), India in carrying out this research work is also gratefully acknowledged.

## Appendix

The Nyquist sampling criterion state that, to sufficiently reproduced a signal with certain frequencies, it should be sampled at a rate at least twice of the maximum frequency contained in that signal. It was named after great American electronic engineer Henry Nyquist. The Nyquist frequency of a signal is the half of the recorded sampling rate in the time domain. So if a signal is sampled at $F_s$ samples/sec rate, its Nyquist frequency simply will be $0.5F_s$ Hz. It is the maximum frequency that can be determined from a discretely sampled data in the time domain. Beyond the Nyquist frequency, *aliasing* will take place which is an effect that causes higher frequencies to be indistinguishable from their low-frequency counterpart. The symmetry about the Nyquist frequency is called folding. Similarly, in the case of a wavenumber, the Nyquist wavenumber will be half of the spatial sampling distance. So, if a signal is captured in the distance domain with $\Delta x$ is the spacing between two sensors, its Nyquist wavenumber will be $0.5 \times \frac{1}{\Delta x}$. As $c = \frac{\omega}{\kappa}$, to obtain part of the dispersion graph with low velocity (*c*) for a particular frequency, one needs to very high wavenumber i.e., low spacing ($\Delta x$) between two sensors. But for a particular test setup spacing between two sensors is fixed, thus we left with certain minimum phase velocity value beyond which no data can be retrieved. As frequency increases, to calculate the same phase velocity value one needs higher and higher wavenumbers, thus the line appears to be inclined with frequencies.

Naskar, T., Kumar J., 2017. Predominant modes for Rayleigh wave propagation using dynamic stiffness matrix approach. Journal of Geophysics and Engineering 14(5), 1032

Naskar, T., Kumar J., 2018. A faster scheme to generate multimodal dispersion plots for Rayleigh wave. Soil Dynamics and Earthquake Engineering Elsevier 117, 280-287.

Nazarian, S., Desai, M.R., 1993. Automated surface wave method: field testing. Journal of Geotechnical Engineering ASCE 119 (7), 1094–1111.

Park, C.B., Miller, R.D., Xia, J., 1998. Imaging dispersion curves of surface waves on multi-channel record. Society of Exploration of Geophysics, 68th Annual Meeting, New Orleans, Louisiana; 1377-1380.

Park, C.B., Miller, R.D., Xia J., 1999. Multichannel analysis of surface waves. Geophysics 64(3), 800-8.

Park, C.B., Miller, R.D., 2008. Roadside passive multichannel analysis of surface waves (MASW). Journal of Environmental and Engineering Geophysics 13(1), 1-11.

Strobbia, C., 2003. Surface wave methods: Acquisition, processing and inversion. Dottorato di Ricerca in Geoingegneria Ambientale, Politecnico Di Torino.

Tokimatsu, K., Tamura, S., Kojima, H., 1992. Effect of multiple modes on Rayleigh wave dispersion characteristics. Journal of Geotechnical Engineering ASCE 118(10), 1529-1543.

Yilmaz, O., 1987. Seismic data processing. Society of Exploration Geophysicists, Tulsa Oklahoma; 526 pp.

Zomorodian, S.M.A., Hunaidi, O., 2006. Inversion of SASW dispersion curves based on maximum flexibility coefficients in the wave number domain. Soil Dynamics and Earthquake Engineering, Elsevier 26, 735–752.**Figure Captions**

Fig. 1. (a) Input subsurface layer properties; and (b) corresponding model dispersion image for all modes.

Fig. 2. A comparison of dispersion plots for synthetic data using 1000 channel by (a) $\omega$-c transform; (b) Zoomed view for $\omega$-c transform; (c) $f$-$\kappa$ transform; and (d) $\tau$-$p$ transform.

Fig. 3. The variation of normalized amplitude with frequency for first three modes of model dispersion curve.

Fig. 4. Dispersion plots by $\omega$-c method for synthetic data using 46 m spread length on the basis of (a) 24 channels; (b) 48 channels; and (c) 96 channels.

Fig. 5. Dispersion plots by $\omega$-$\kappa$ method for synthetic data using 46 m spread length and $\Lambda=50$ zero padding on the basis of (a) 24 channels; (b) 48 channels; and (c) 96 channels.

Fig. 6. Dispersion plots by $\tau$-$p$ method for synthetic data using 46 m spread length on the basis of (a) 24 channels; (b) 48 channels; and (c) 96 channels.

Fig. 7. Dispersion plots by $\omega$-c method for synthetic data using 92 m spread length on the basis of (a) 24 channels; (b) 48 channels; and (c) 96 channels.

Fig. 8. Dispersion plots by $\omega$-$\kappa$ method for synthetic data using 92 m spread length and $\Lambda=50$ zero padding on the basis of (a) 24 channels; (b) 48 channels; and (c) 96 channels.

Fig. 9. Dispersion plots by $\tau$-$p$ method for synthetic data using 92 m spread length on the basis of (a) 24 channels; (b) 48 channels; and (c) 96 channels.

Fig. 10. Dispersion plots by $\omega$-c method for synthetic data using 184 m spread length on the basis of (a) 24 channels; (b) 48 channels; and (c) 96 channels.

Fig. 11. Dispersion plots by $\omega$-$\kappa$ method for synthetic data using 184 m spread length and $\Lambda=50$ zero padding on the basis of (a) 24 channels; (b) 48 channels; and (c) 96 channels.

Fig. 12. Dispersion plots by $\tau$-$p$ method for synthetic data using 184 m spread length on the basis of (a) 24 channels; (b) 48 channels; and (c) 96 channels.

Fig. 13. Input signal for different offsets in a time domain for synthetic data with spread length (X) of 184 m (a) for 24 channels; (b) for 96 channels; (c) for 24 channels with smooth spline interpolation to 96 channels.

Fig. 14. Dispersion plots for synthetic data using 184 m spread length on the basis of (a) 24 channels; (b) 96 channels; and (c) 24 channel, interpolated to 96 channel.

Fig. 15. A comparison of dispersion images for synthetic data by using 96 channels on the basis of $\tau$-$p$ transform for (a) $F_s = 5$ kHz; (b) $F_s = 10$ kHz; (c) $F_s = 25$ kHz; and (d) $F_s = 100$ kHz.

Fig. 16. A comparison of two-dimensional dispersion plots for synthetic data by using 96 channels on the basis of $\omega$-$\kappa$ transform for different data zero padding with (a) $\Lambda=1$; (b) $\Lambda=5$; (c) $\Lambda=25$; and (d) $\Lambda=50$.

Fig. 17. Dispersion plots for site-1 using 48 channels on the basis of (a) $\omega$-$c$ transform; (b) $f$-$\kappa$ transform; (c) $\tau$-$p$ transform; and (d) theoretical profile from inverse analysis.

Fig. 18. Two-dimensional dispersion plots for field data using 48 channels on the basis of (a) $\omega$-$c$ transform; (b) $f$-$\kappa$ transform; (c) $\tau$-$p$ transform; and (d) all the three methods.

Fig. 19. (a) A comparison between dispersion image and predicted theoretical curve from inverse analysis; (b) theoretical soil profile from inverse analysis for Site-2

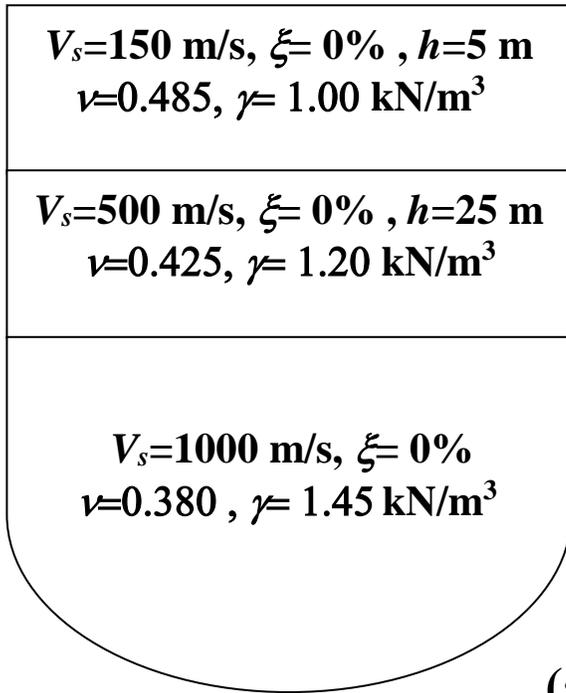
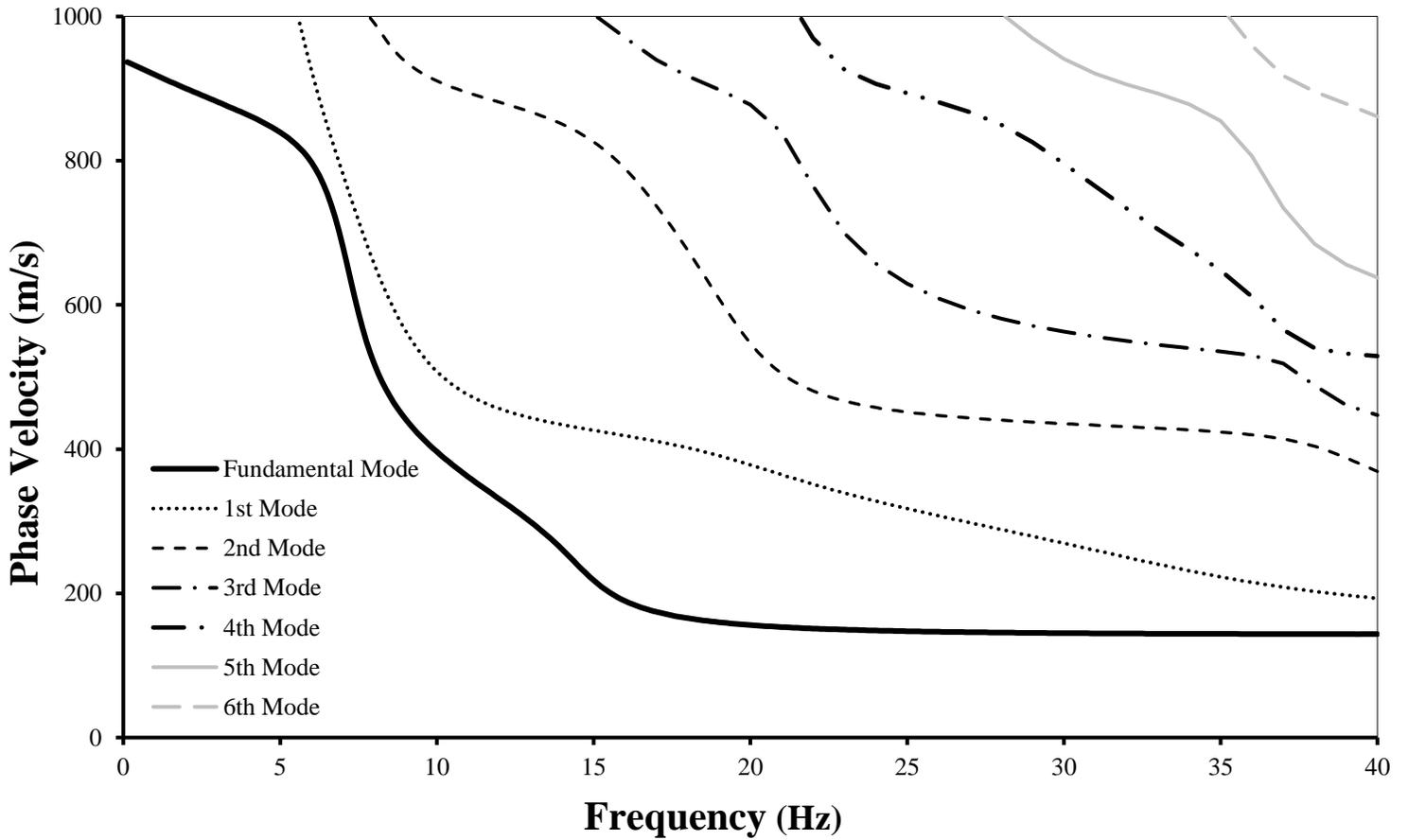

**Figure 1**. (a) Input subsurface layer properties; and (b) corresponding model dispersion image for all modes.

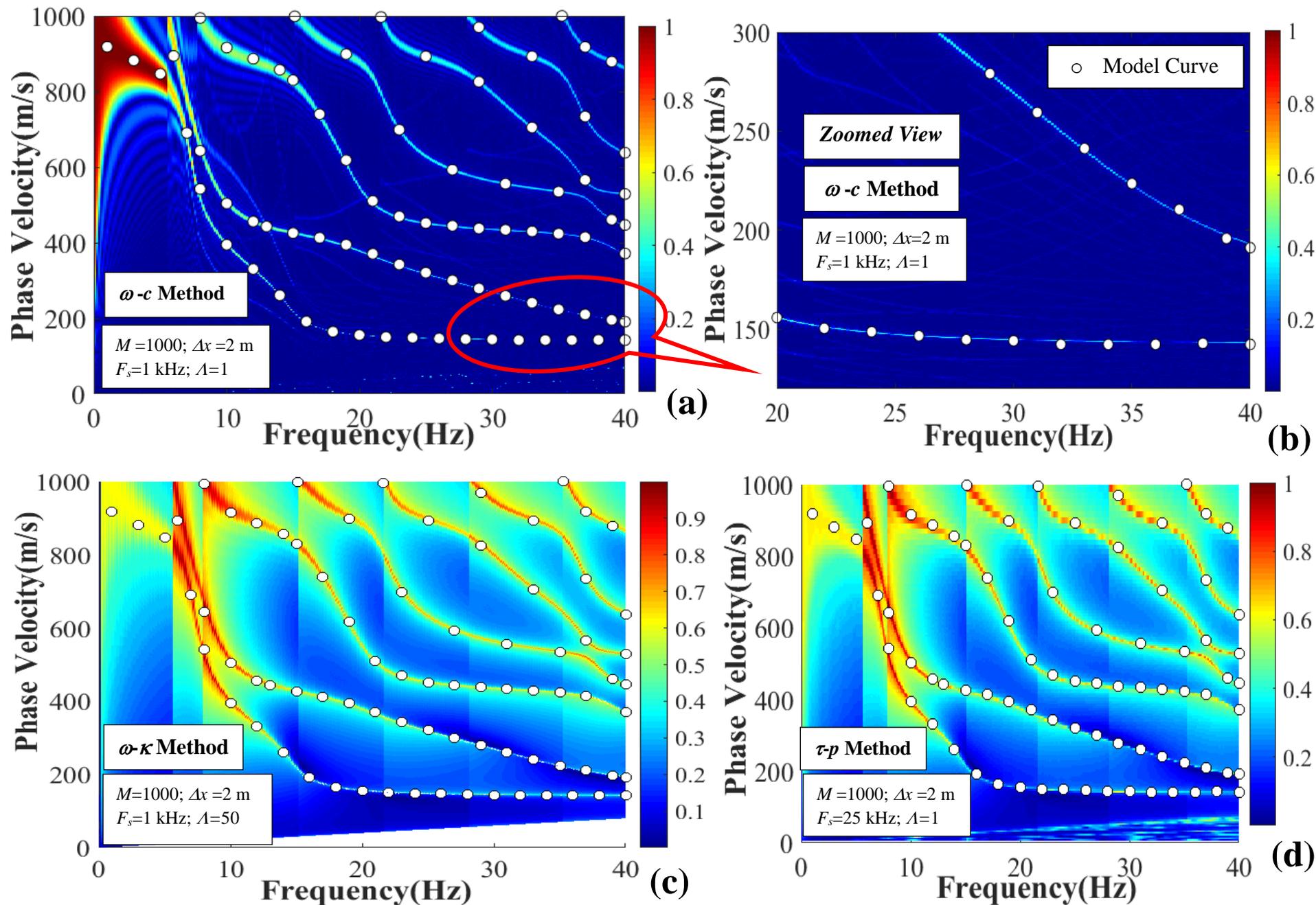

**Figure 2.** A comparison of dispersion plots for synthetic data using 1000 channel by (a) $\omega$-$c$ transform; (b) Zoomed view for $\omega$-$c$ transform; (c) $\omega$-$\kappa$ transform; and (d) $\tau$-$p$ transform.

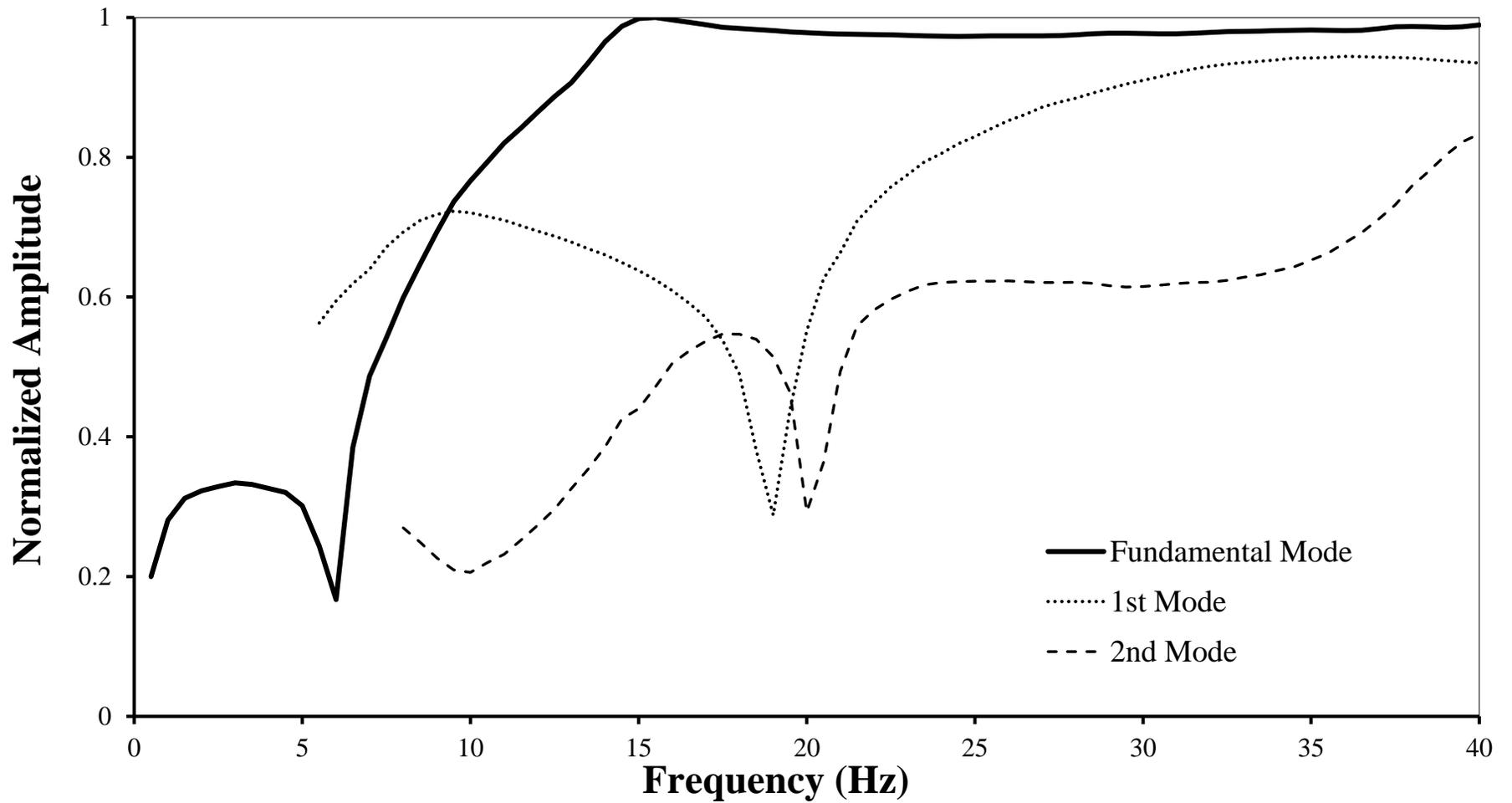

**Figure 3**. The variation of normalized amplitude with frequency for first three modes of model dispersion curve.

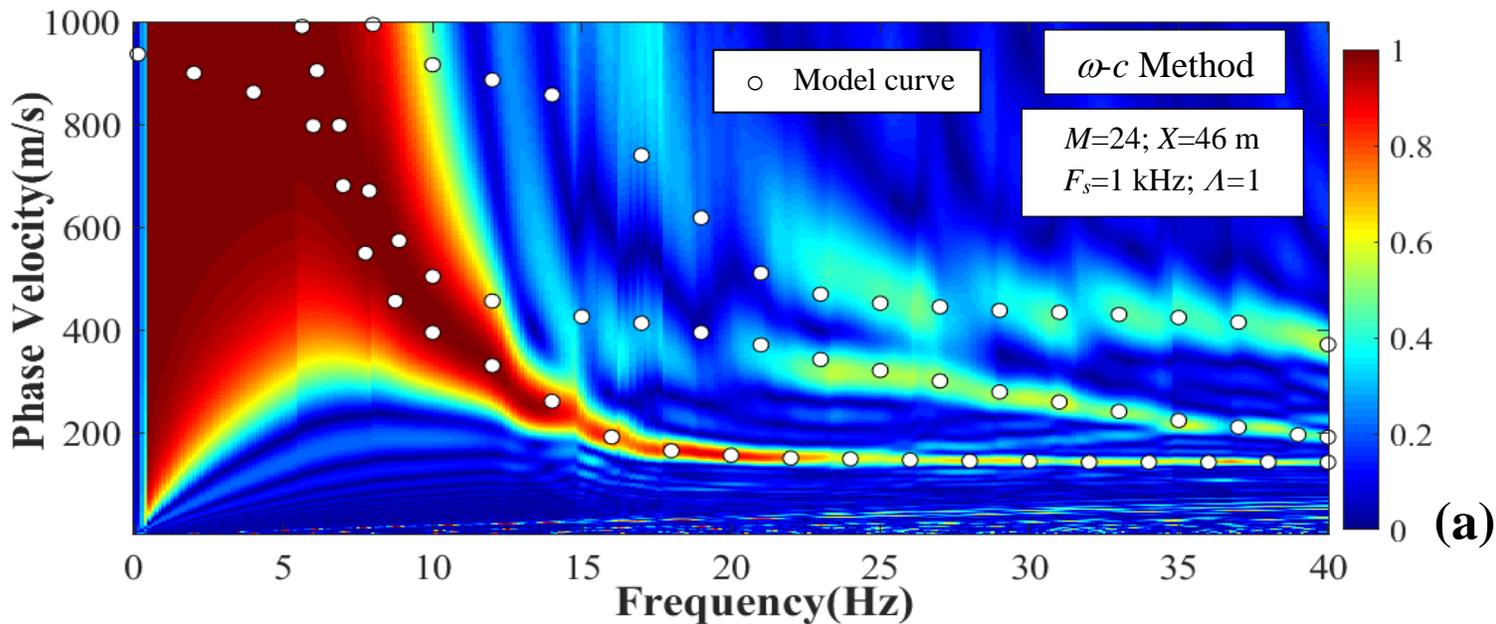
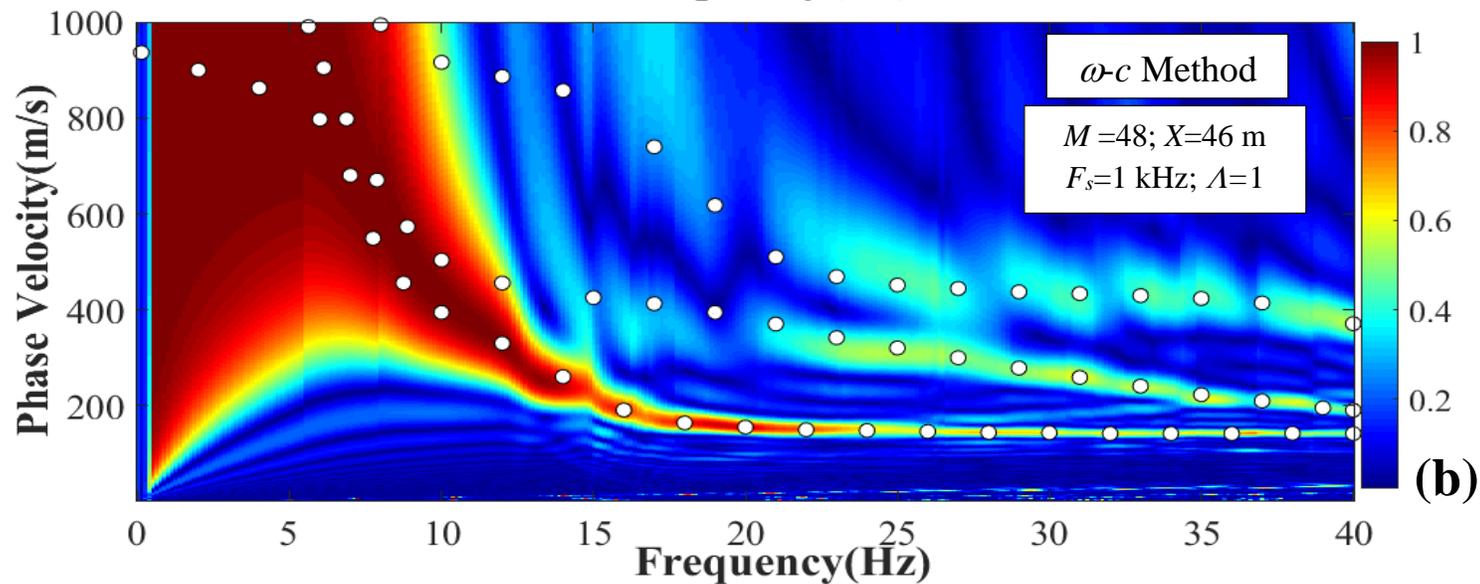
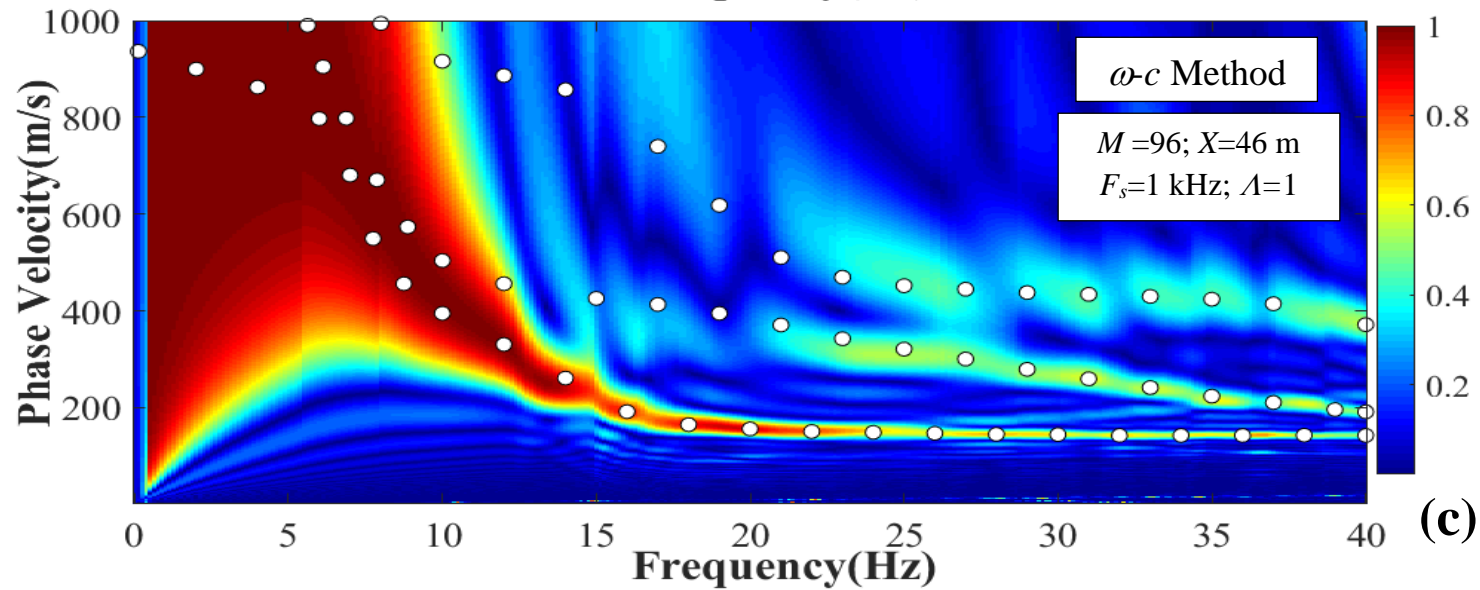

**Figure 4**. Dispersion plots by *ω-c* method for synthetic data using 46 m spread length on the basis of (a) 24 channels; (b) 48 channels; and (c) 96 channels.

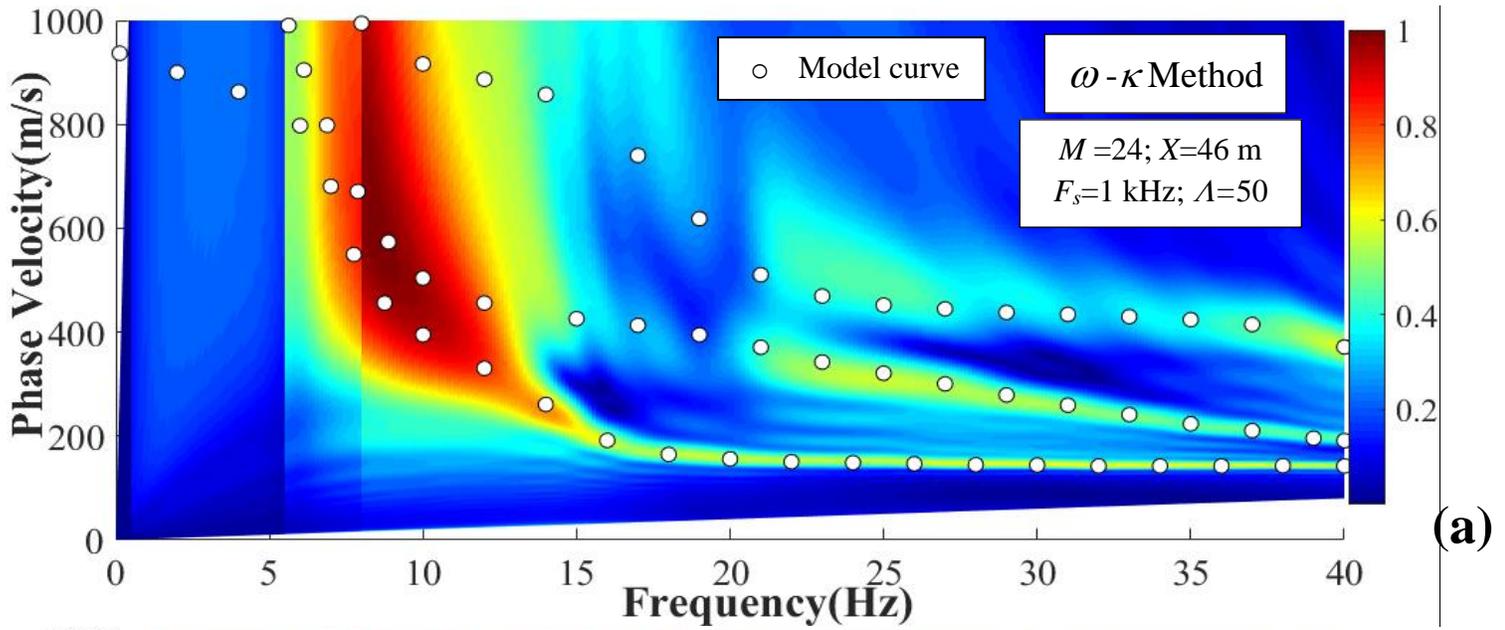
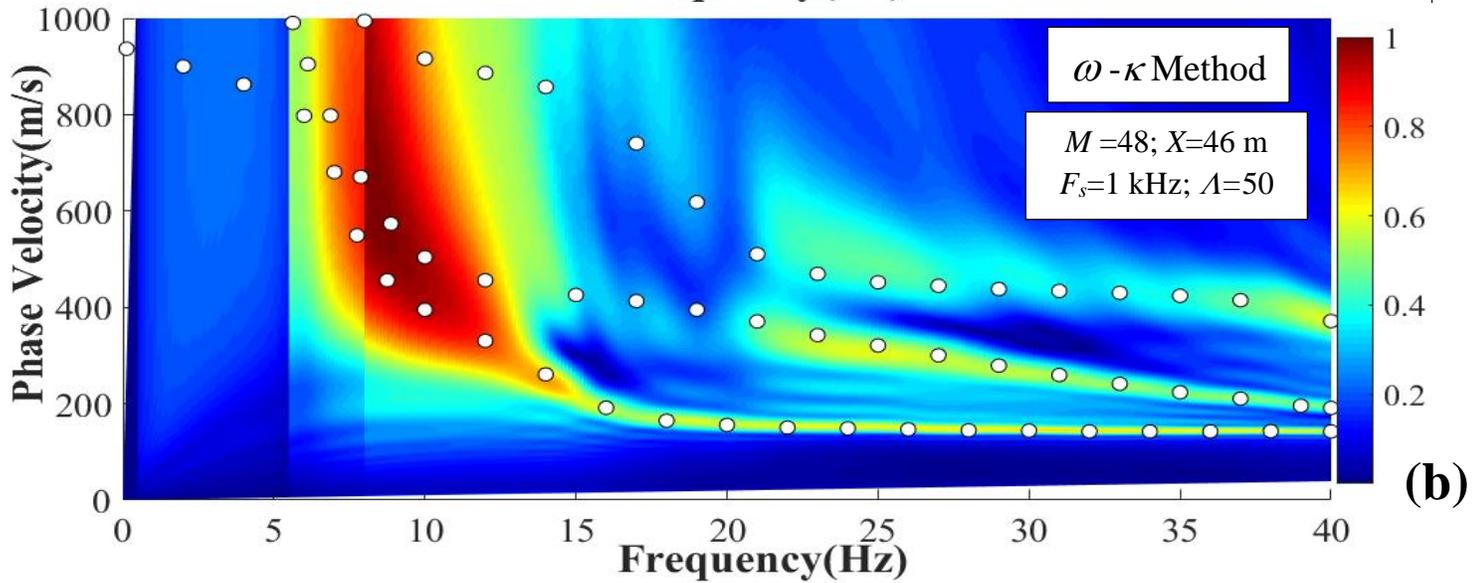
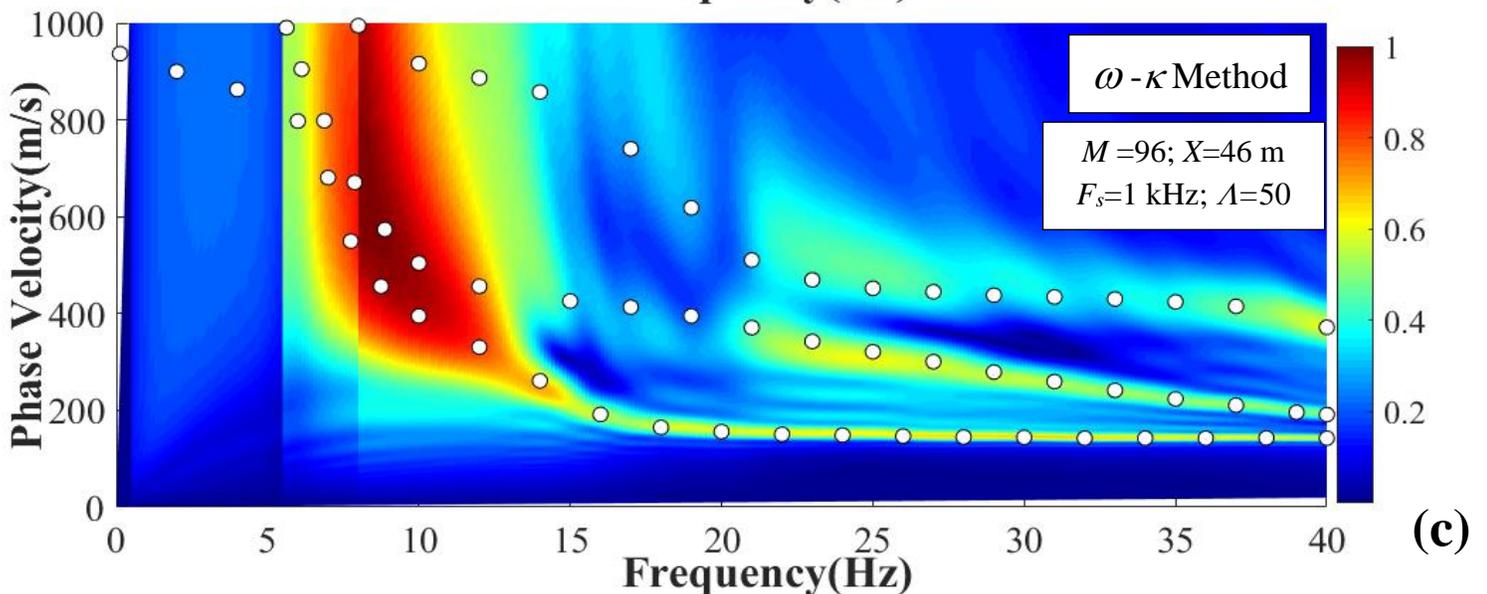

**Figure 5**. Dispersion plots by $\omega$-$\kappa$ method for synthetic data using 46 m spread length and $\Lambda$=50 zero padding on the basis of (a) 24 channels; (b) 48 channels; and (c) 96 channels.

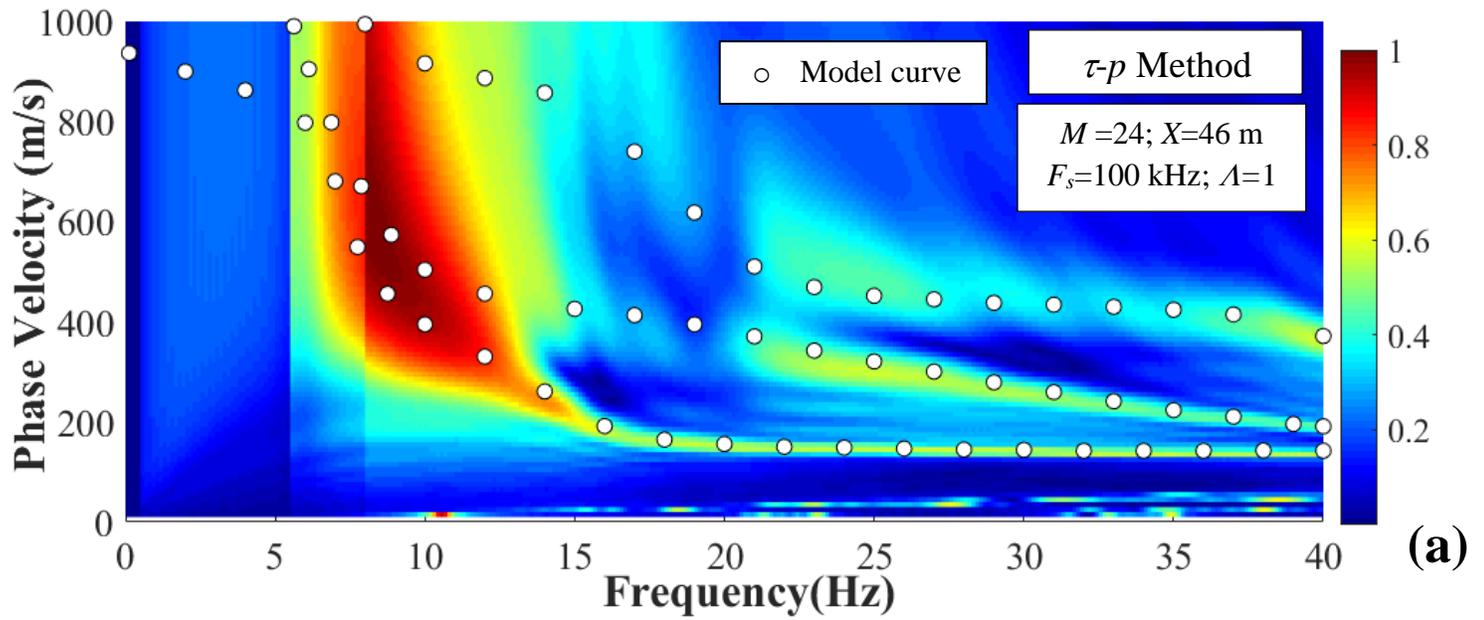
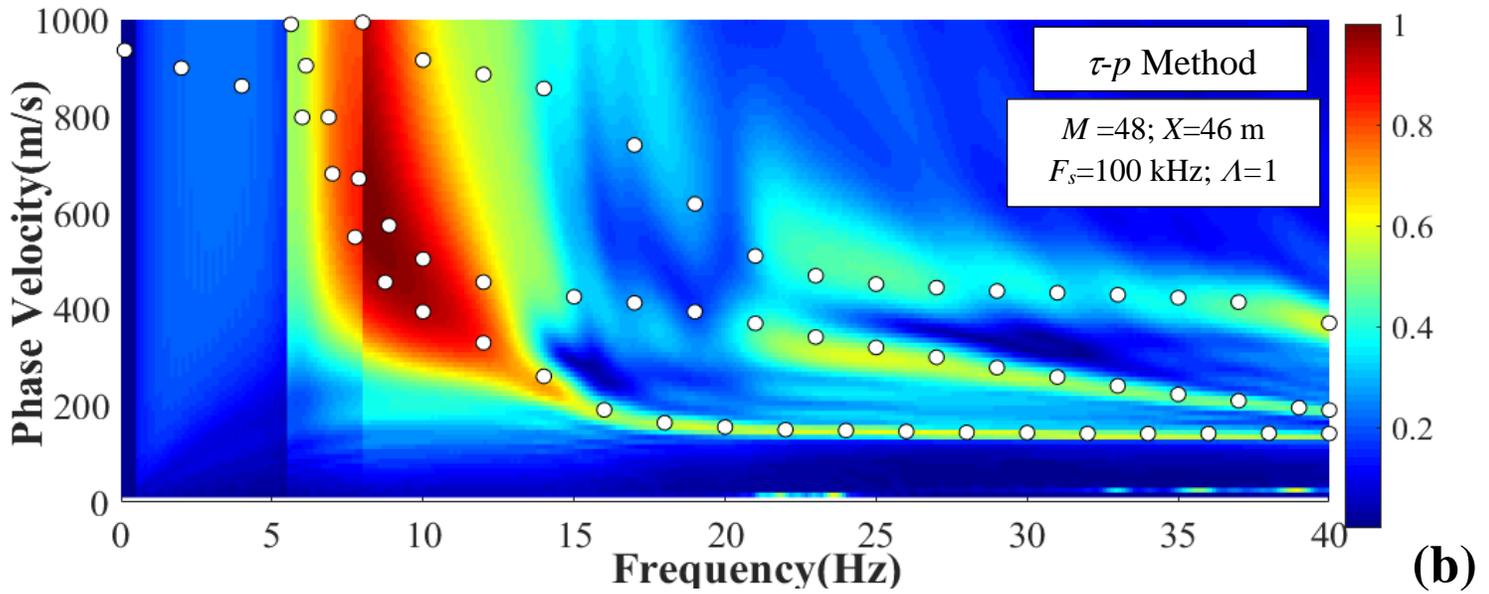
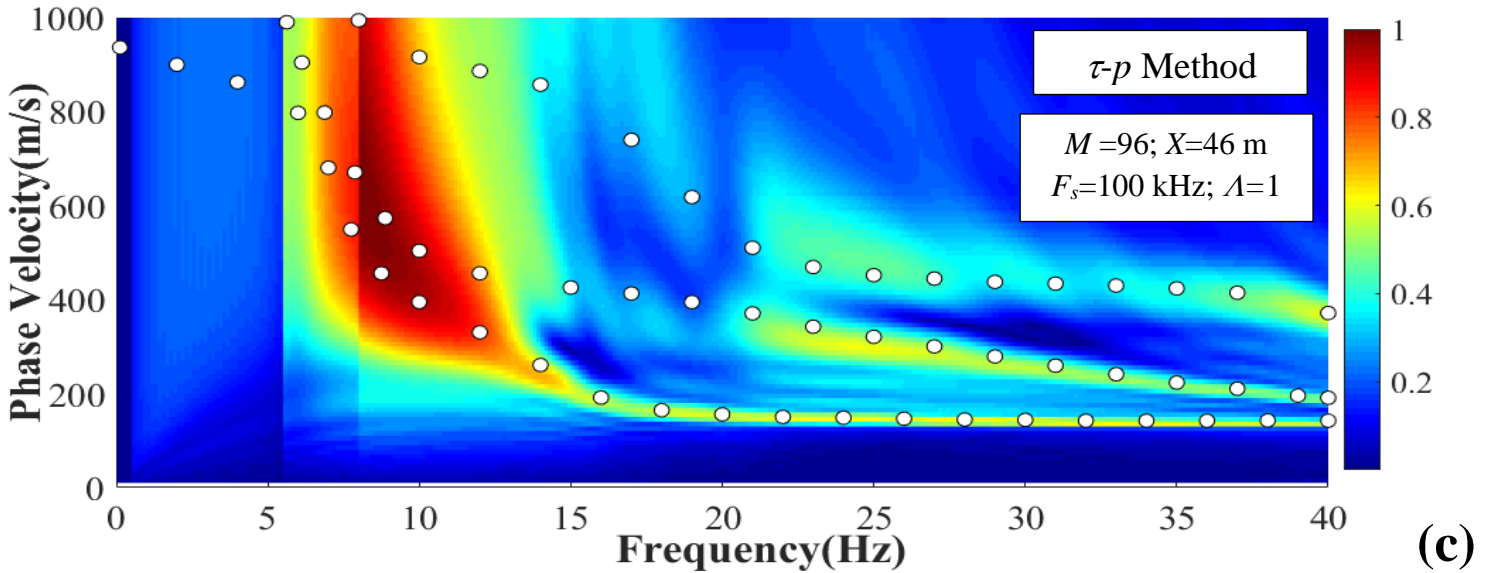

**Figure 6**. Dispersion plots by $\tau$-$p$ method for synthetic data using 46 m spread length on the basis of (a) 24 channels; (b) 48 channels; and (c) 96 channels.

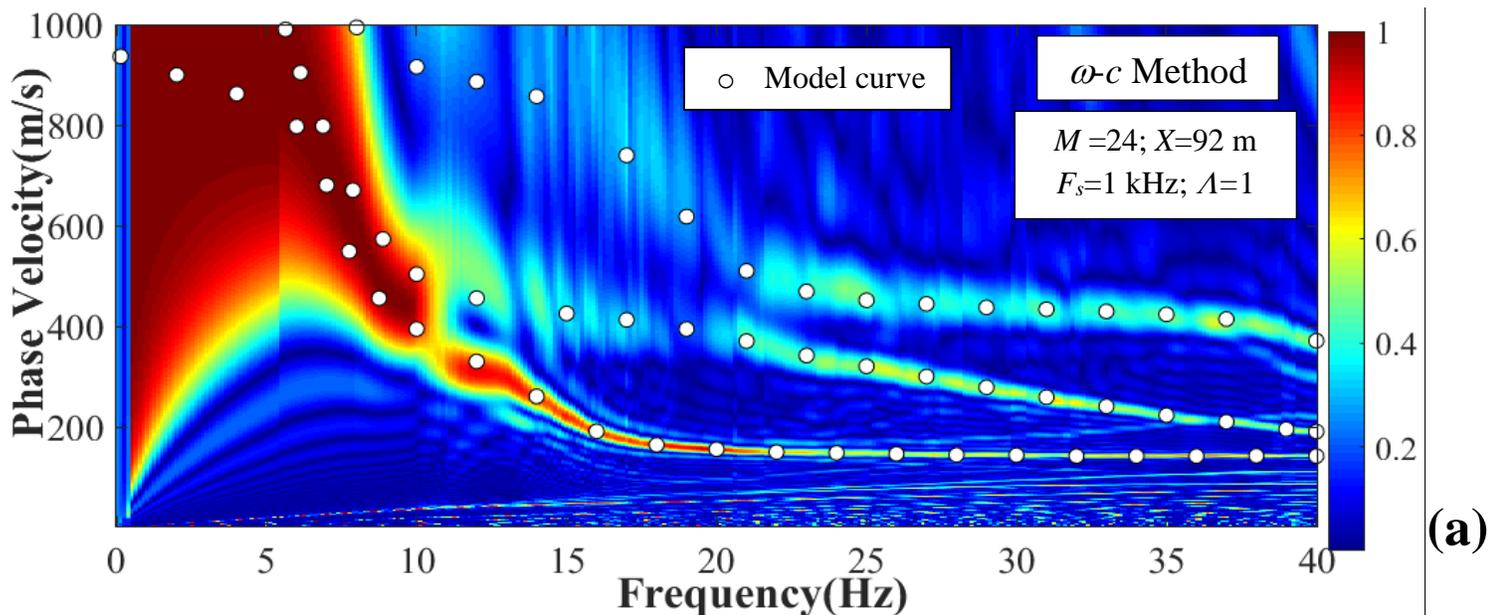
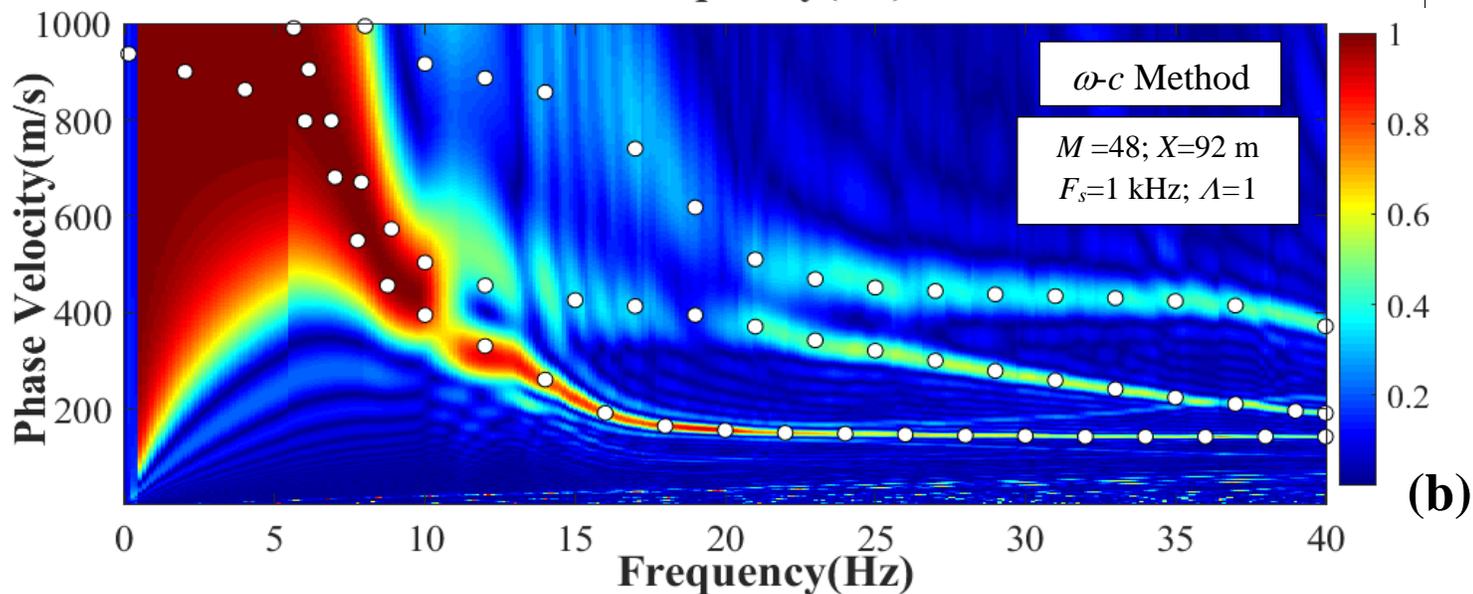
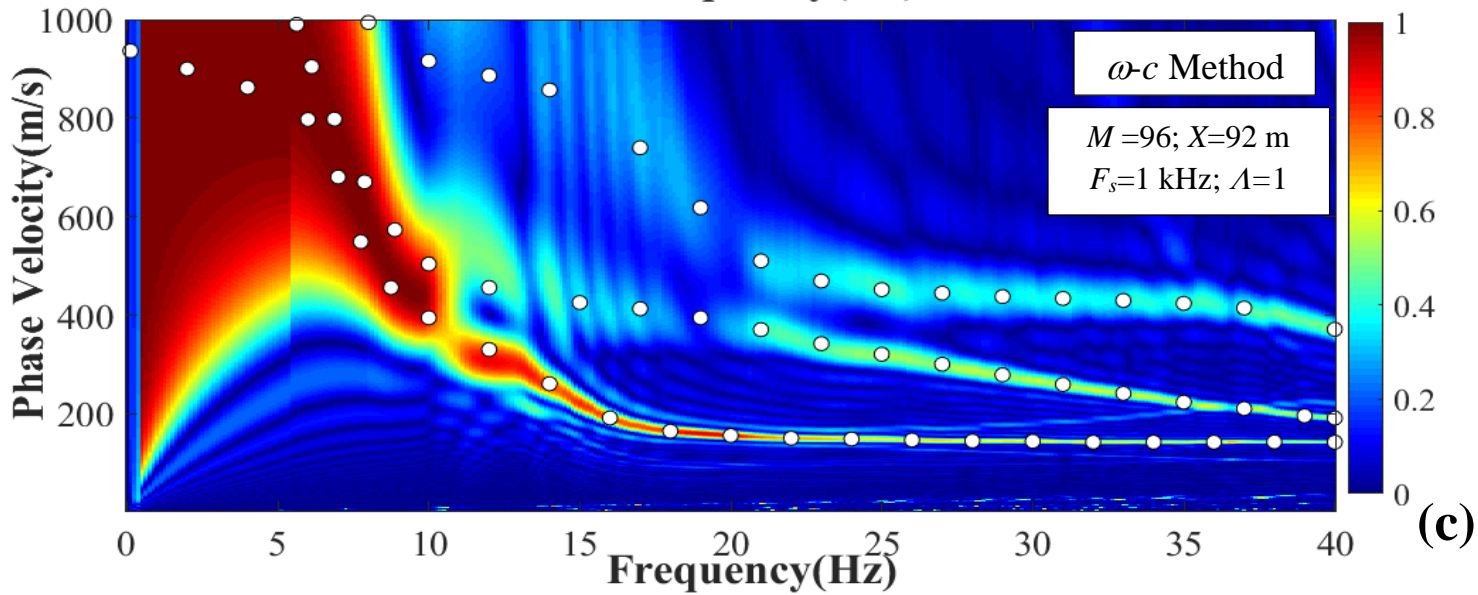

**Figure 7**. Dispersion plots by $\omega$-$c$ method for synthetic data using 92 m spread length on the basis of (a) 24 channels; (b) 48 channels; and (c) 96 channels.

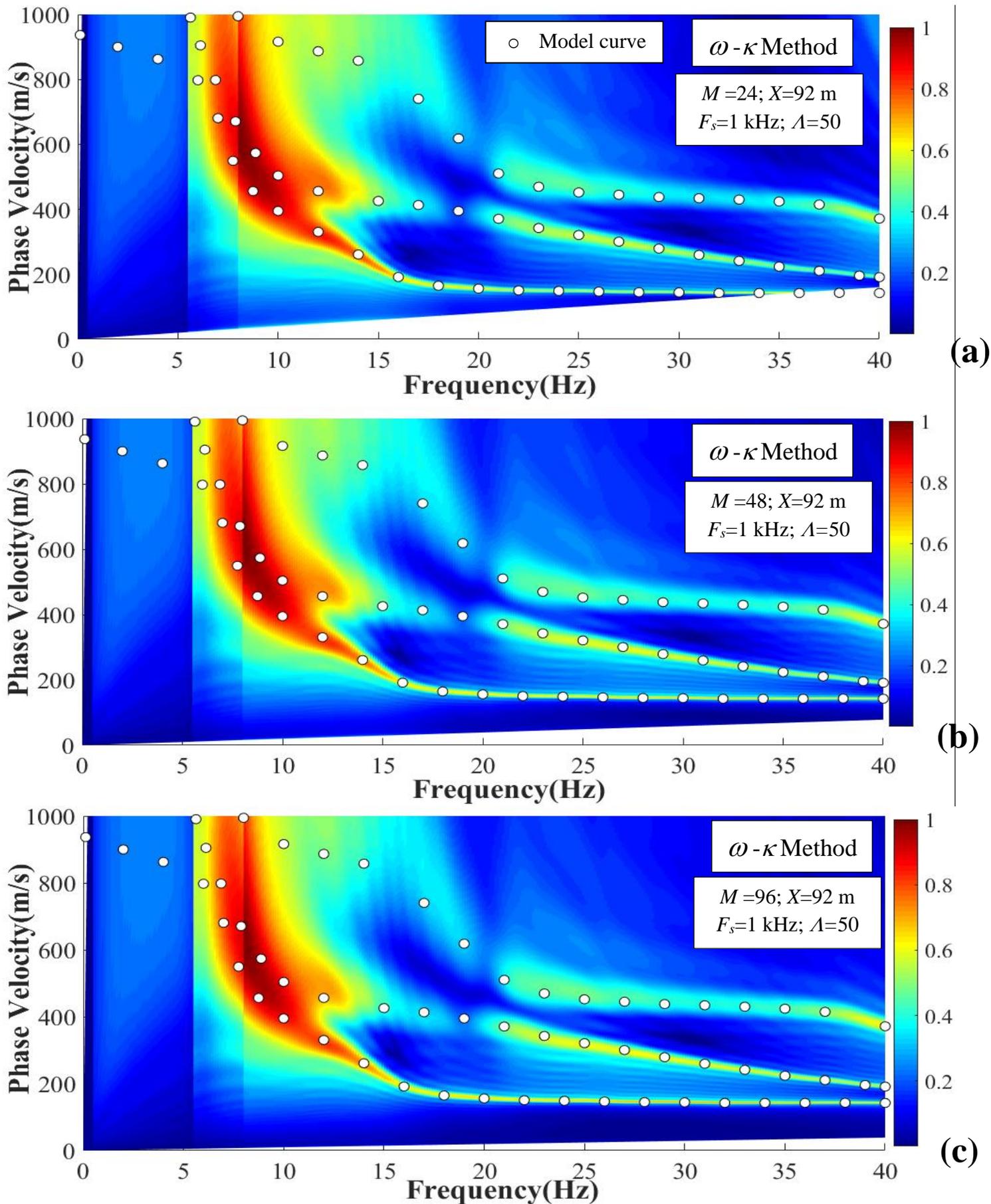

**Figure 8.** Dispersion plots by $\omega$-$\kappa$ method for synthetic data using 92 m spread length and $\Lambda$=50 zero padding on the basis of (a) 24 channels; (b) 48 channels; and (c) 96 channels.

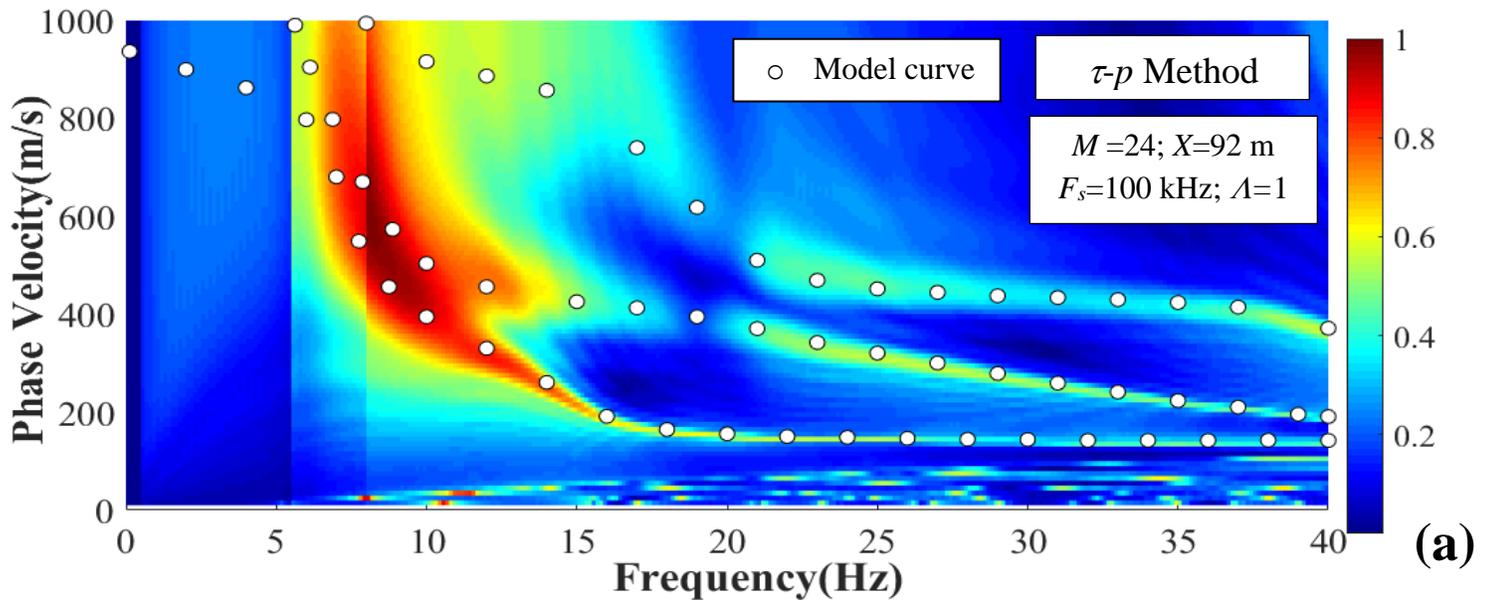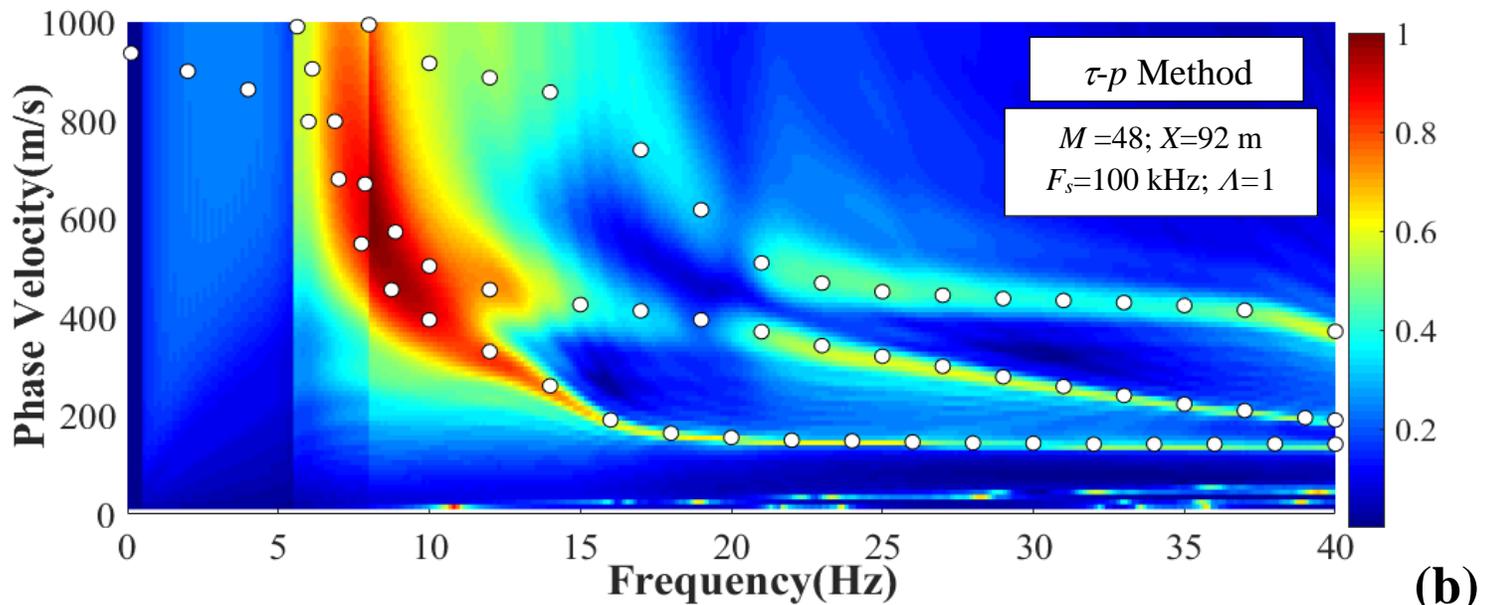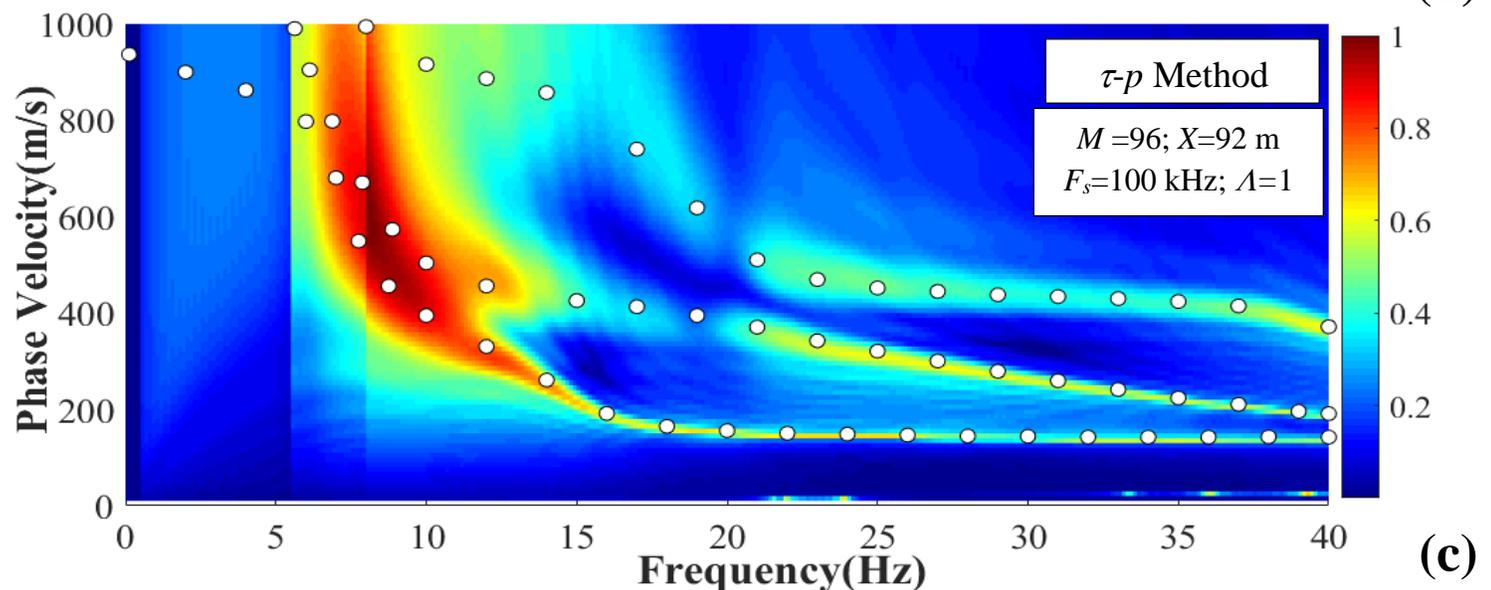

**Figure 9.** Dispersion plots by $\tau$-$p$ method for synthetic data using 92 m spread length on the basis of (a) 24 channels; (b) 48 channels; and (c) 96 channels.

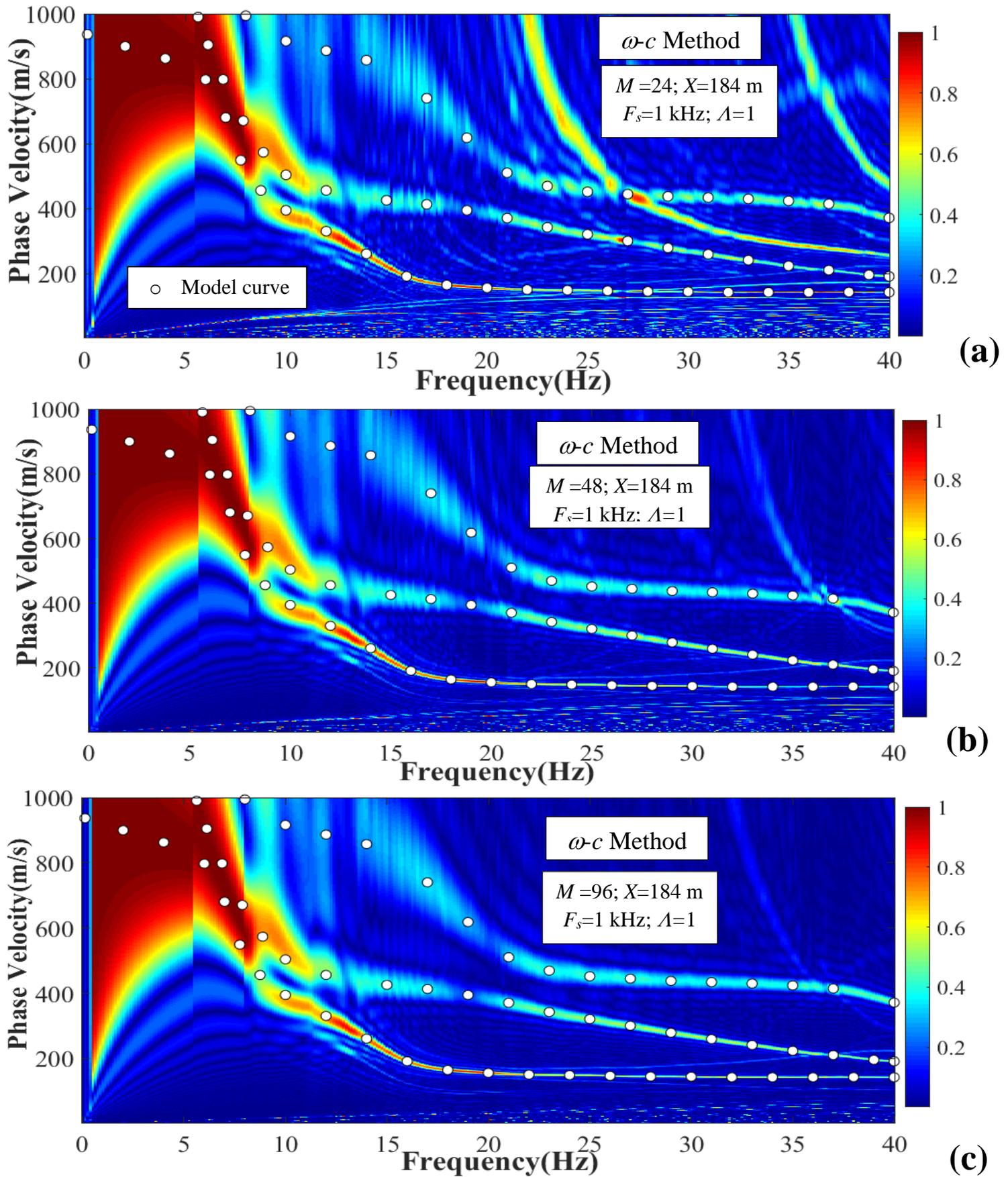

**Figure 10**. Dispersion plots by $\omega$-$c$ method for synthetic data using 184 m spread length on the basis of (a) 24 channels; (b) 48 channels; and (c) 96 channels.

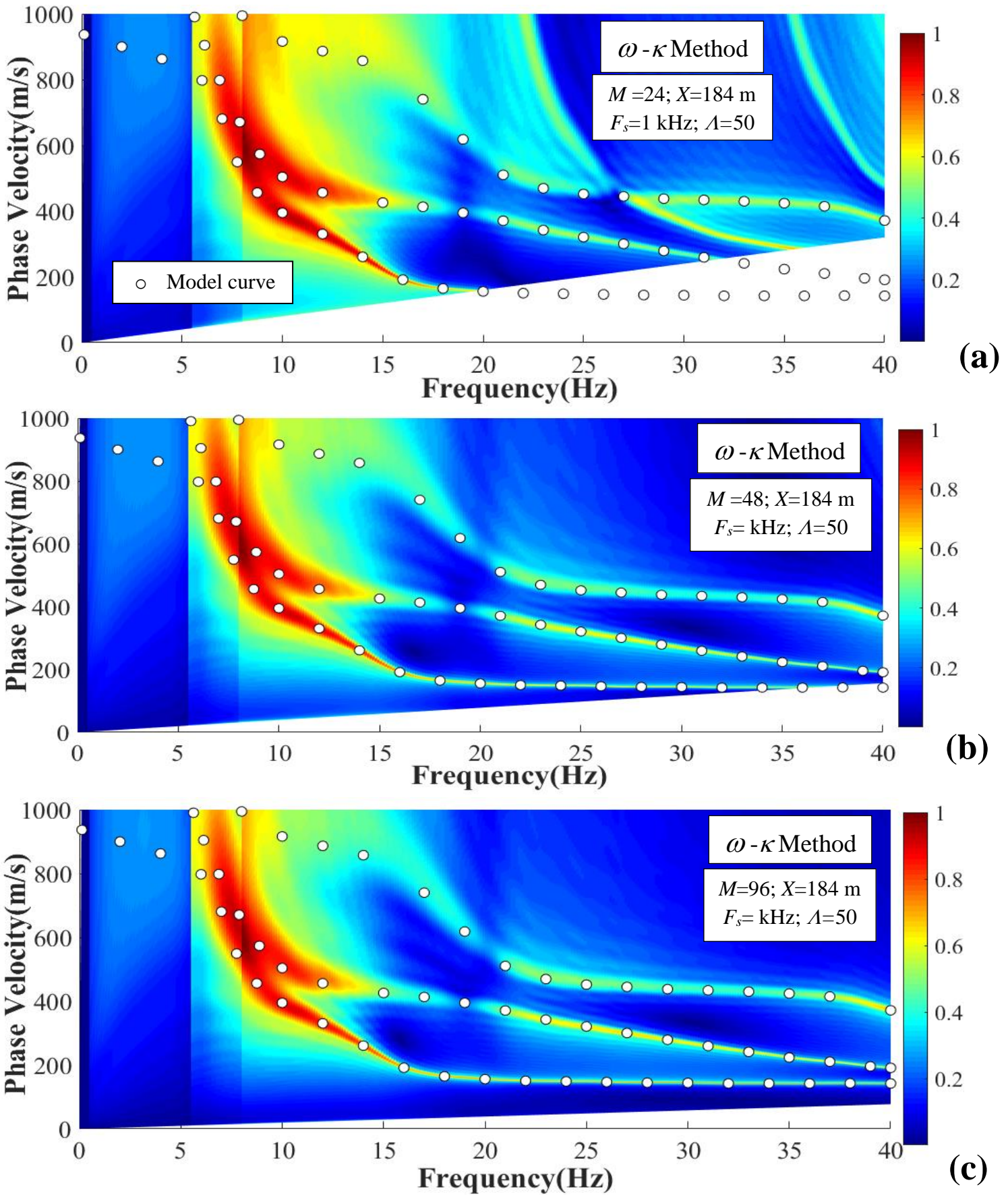

**Figure 11**. Dispersion plots by $\omega$-$\kappa$ method for synthetic data using 184 m spread length and $\Lambda$=50 zero padding on the basis of (a) 24 channels; (b) 48 channels; and (c) 96 channels.

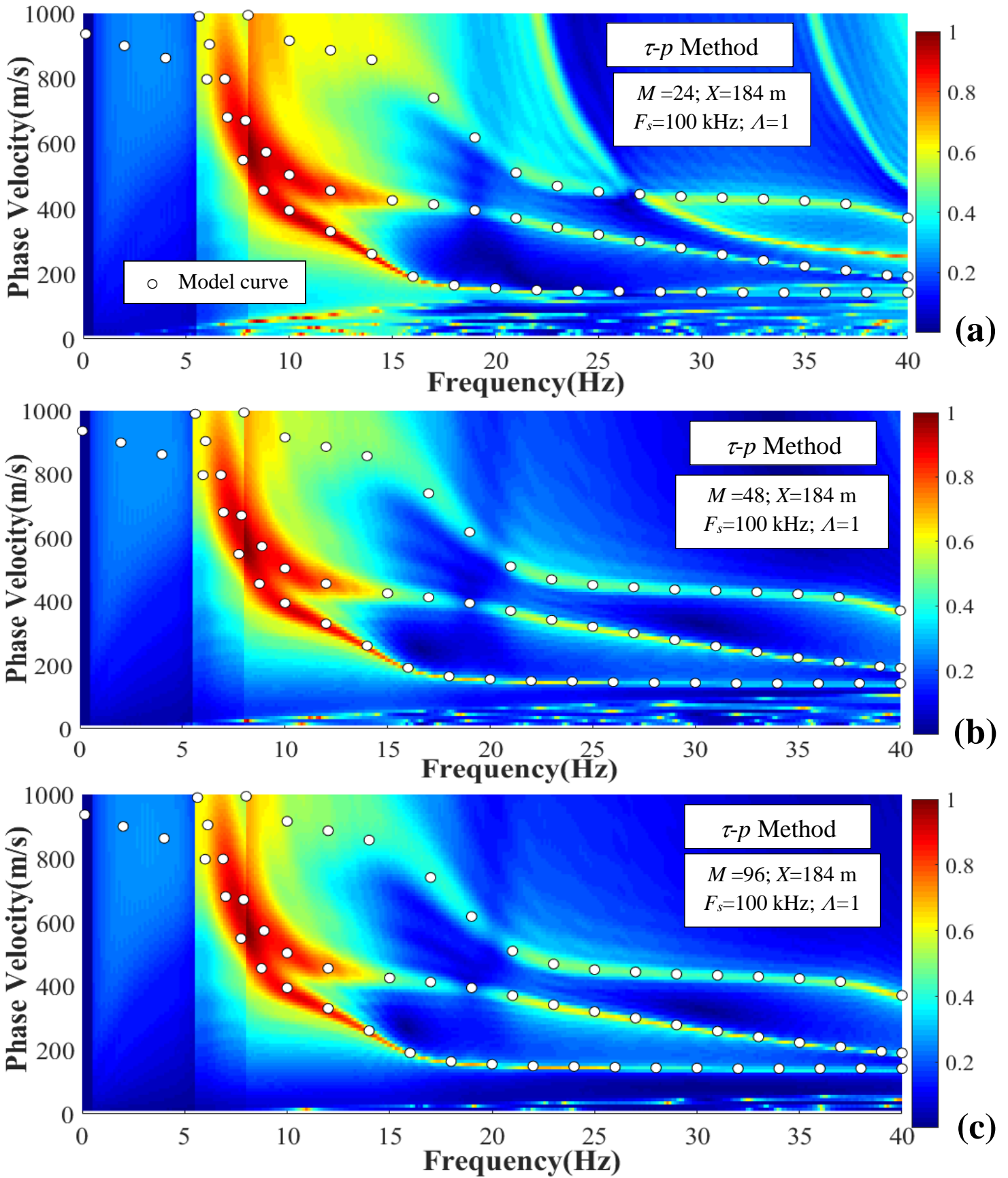

**Figure 12.** Dispersion plots by $\tau\text{-}p$ method for synthetic data using 184 m spread length on the basis of (a) 24 channels; (b) 48 channels; and (c) 96 channels.

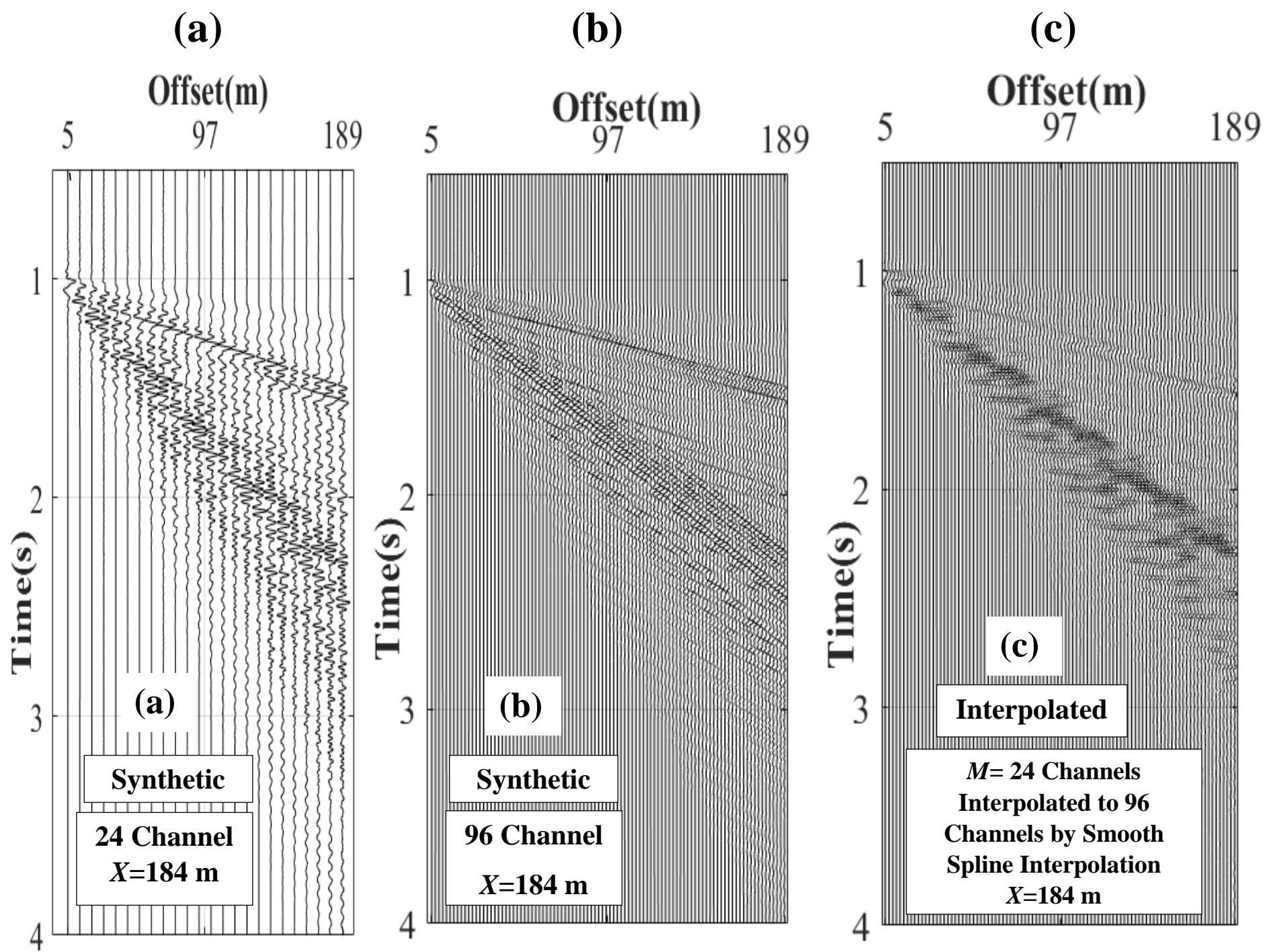

**Figure 13.** Input signal for different offsets in a time domain for synthetic data with spread length ($X$) of 184 m (a) for 24 channels; (b) for 96 channels; (c) for 24 channels with smooth spline interpolation to 96 channels.

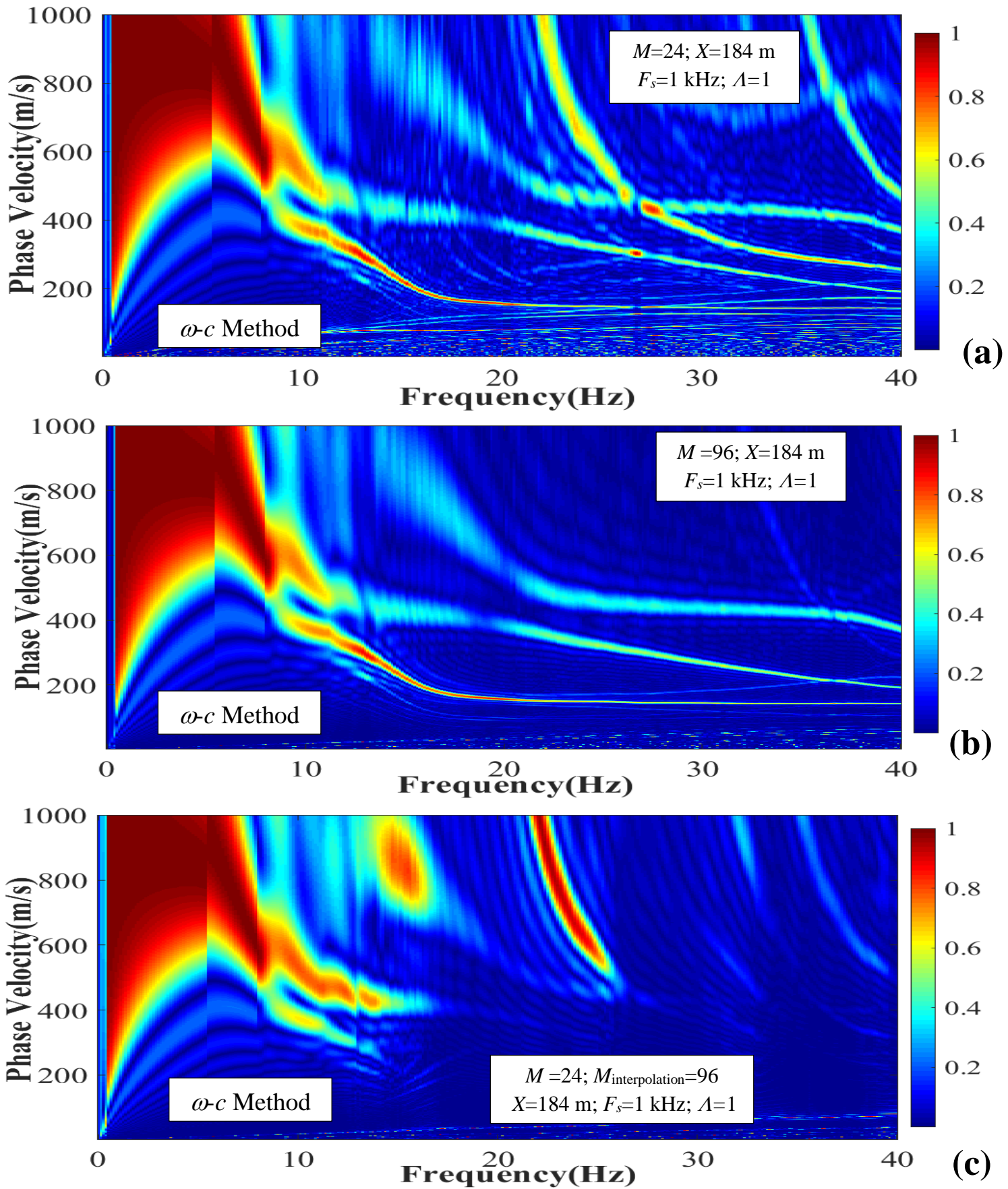

**Figure 14**. Dispersion plots for synthetic data using 184 m spread length on the basis of (a) 24 channels; (b) 96 channels; and (c) 24 channel, interpolated to 96 channel.

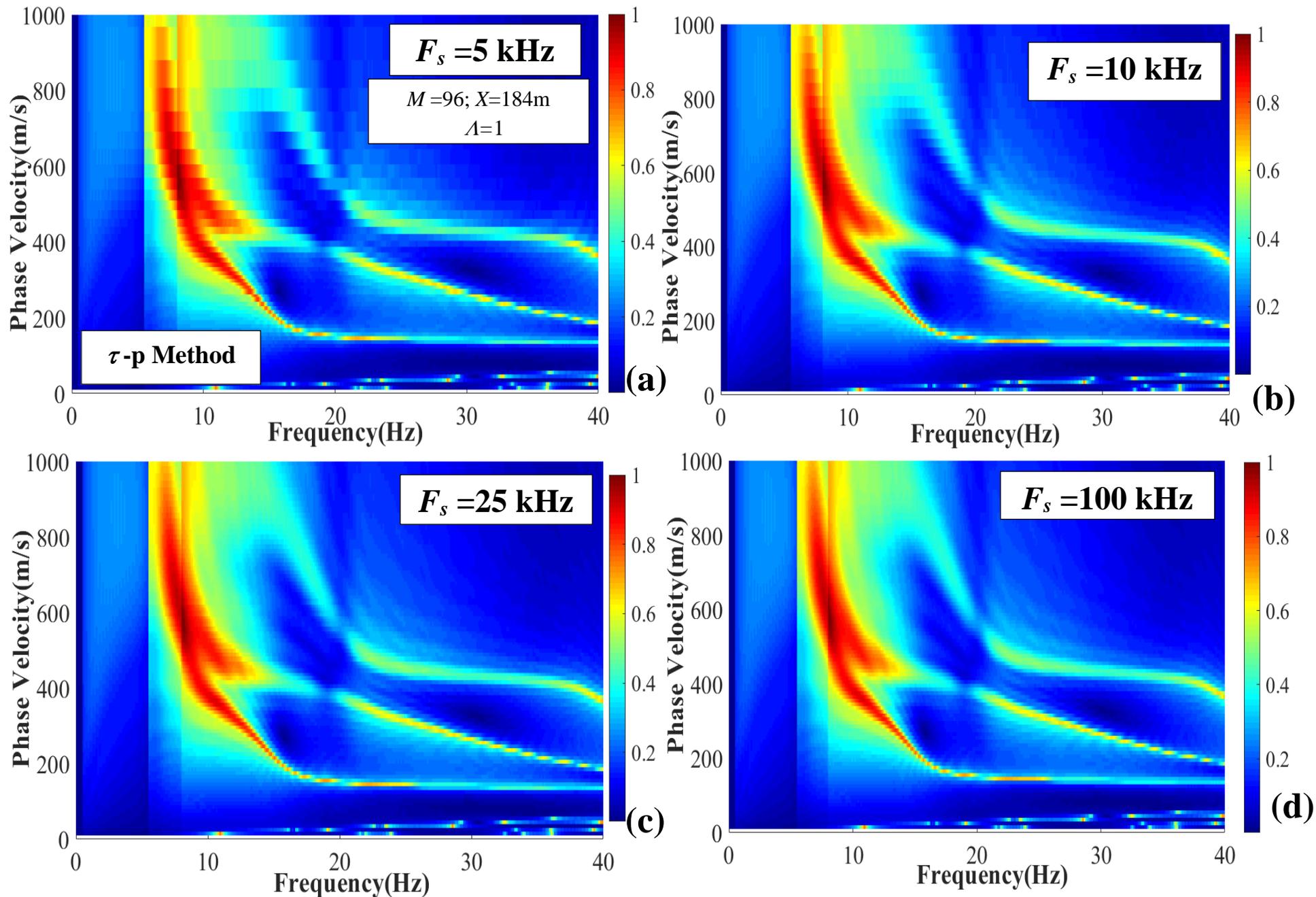

**Figure 15**. A comparison of dispersion images for synthetic data by using 96 channels on the basis of $\tau$-$p$ transform for (a) $F_s$ = 5 kHz; (b) $F_s$ = 10 kHz; (c) $F_s$ = 25 kHz; and (d) $F_s$ = 100 kHz.

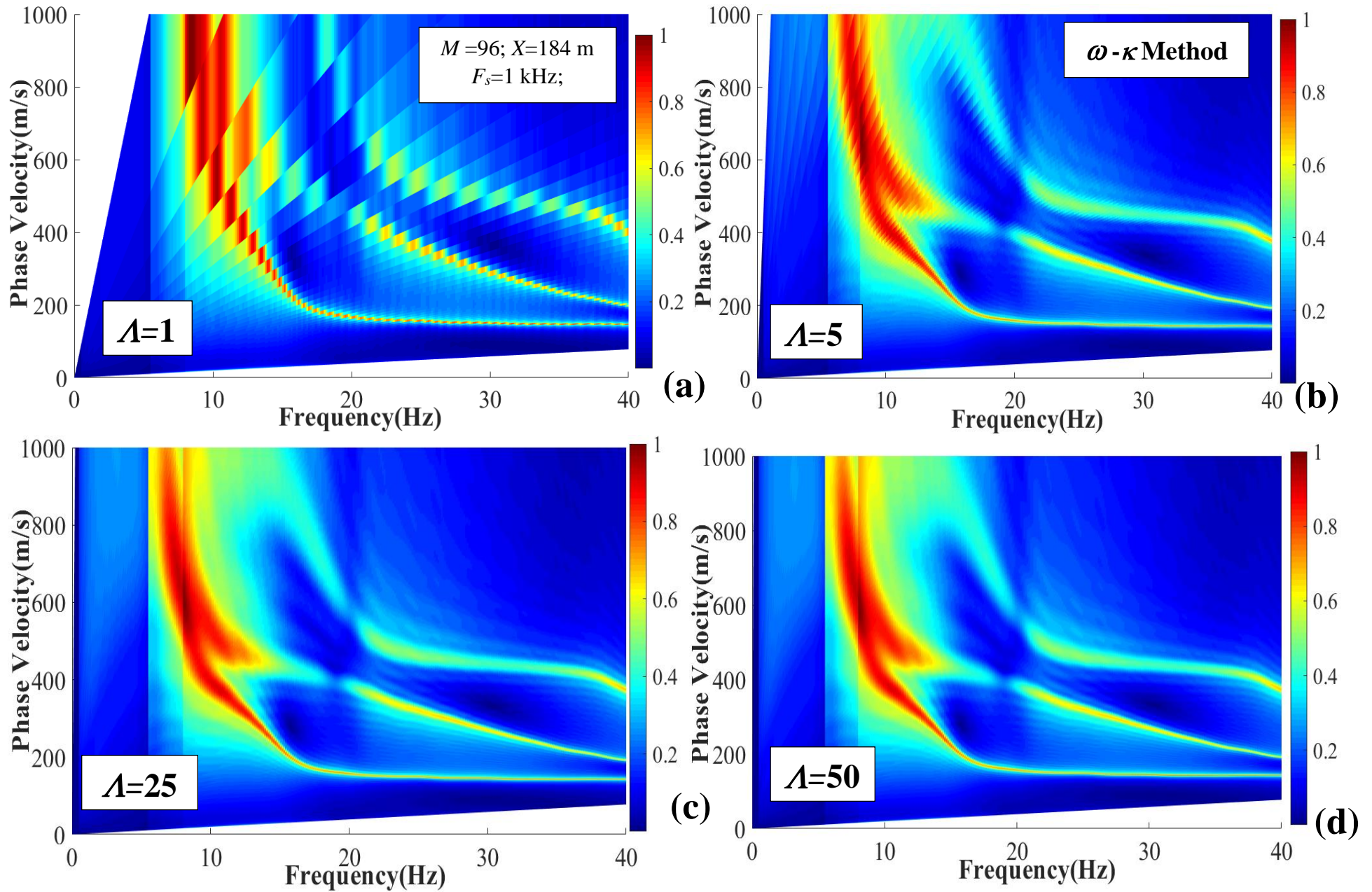

**Figure 16**. A comparison of two-dimensional dispersion plots for synthetic data by using 96 channels on the basis of $\omega$-$\kappa$ transform for different data zero padding with (a) $\Lambda$=1; (b) $\Lambda$=5; (c) $\Lambda$=25; and (d) $\Lambda$=50.

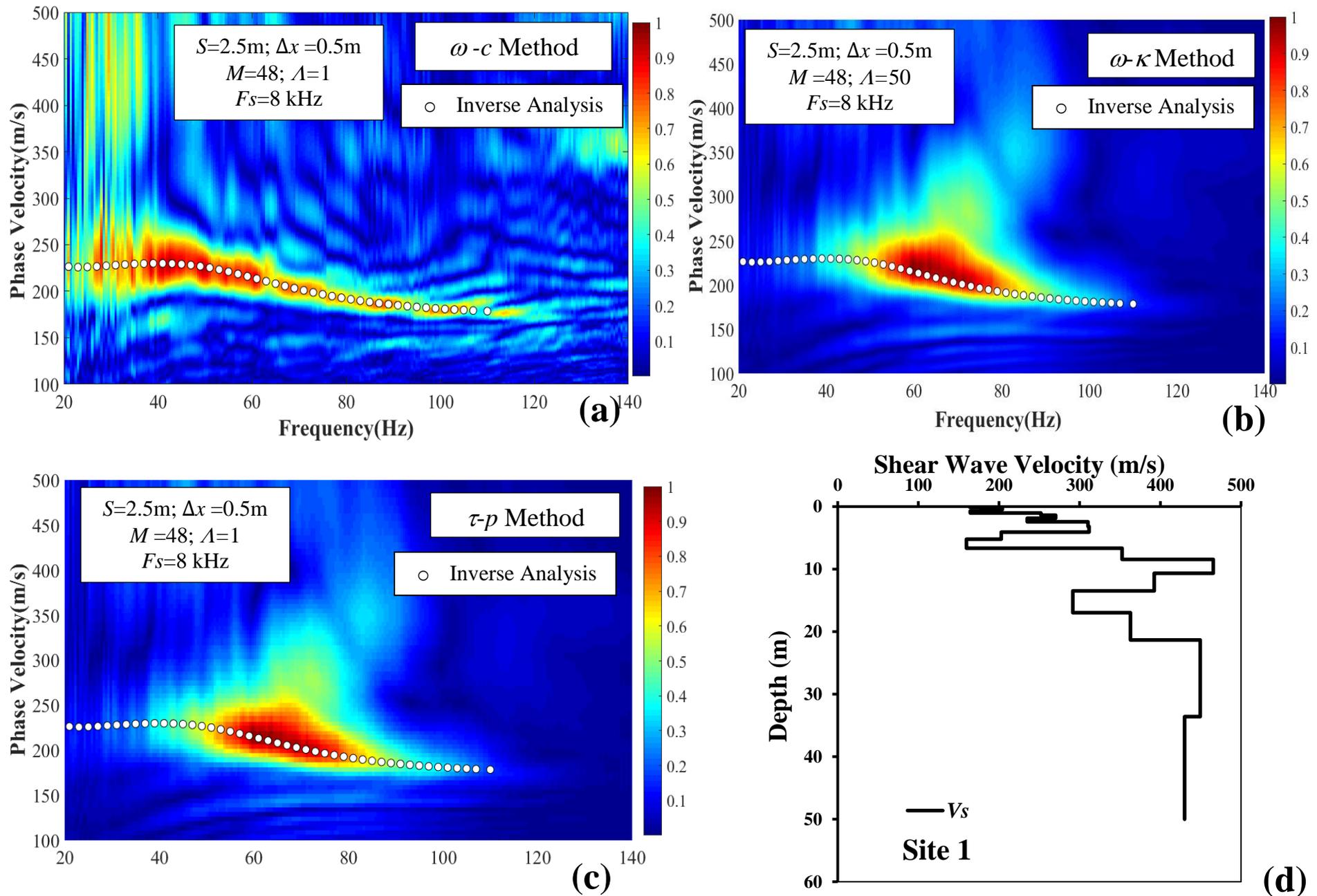

**Figure 17**. Dispersion plots for site-1 using 48 channels on the basis of (a) $\omega$-$c$ transform; (b) $\omega$-$\kappa$ transform; (c) $\tau$-$p$ transform; and (d) theoretical profile from inverse analysis.

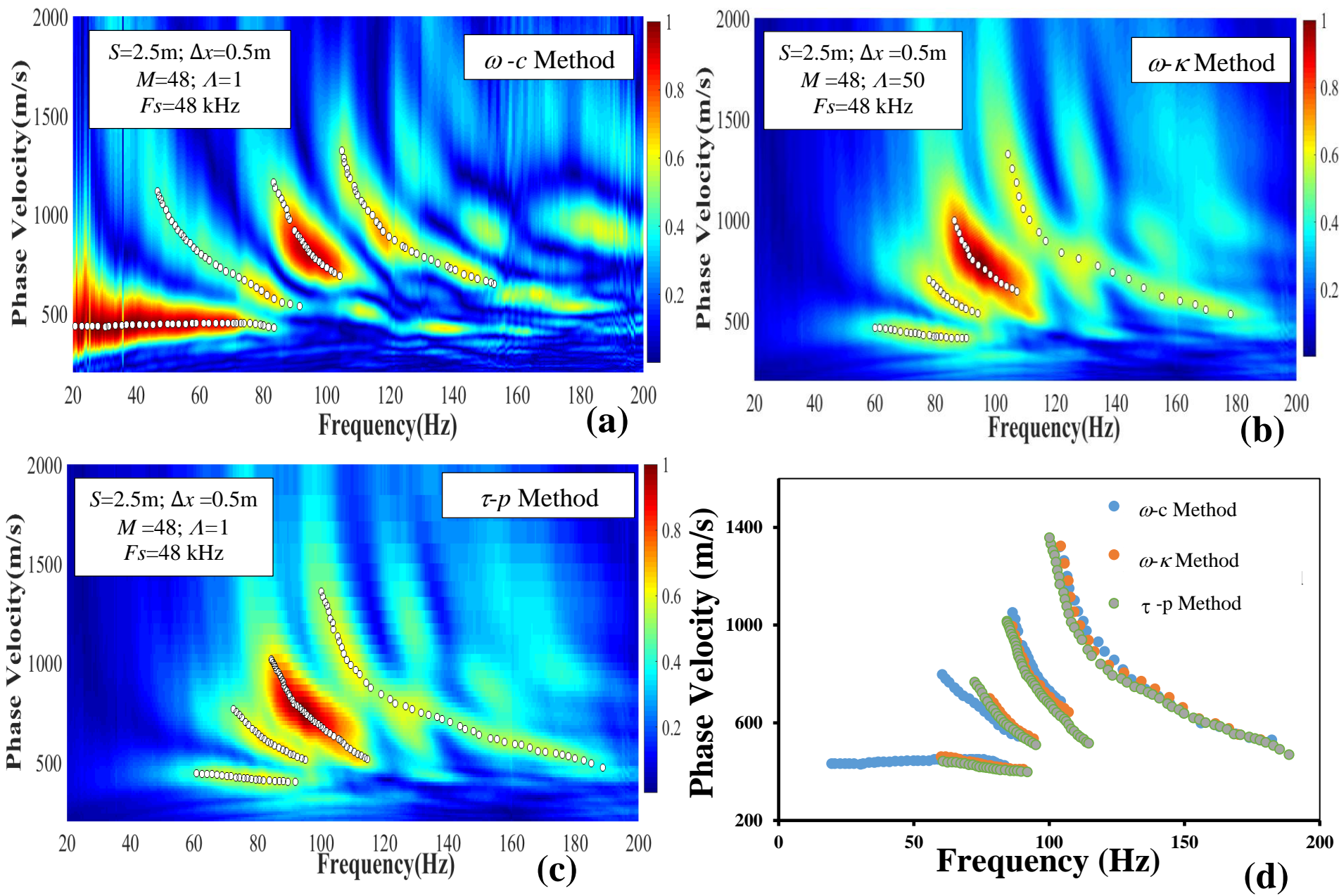

**Figure 18**. Two-dimensional dispersion plots for field data using 48 channels on the basis of (a) $\omega$-$c$ transform; (b) $\omega$-$\kappa$ transform; (c) $\tau$-$p$ transform; and (d) all the three methods.

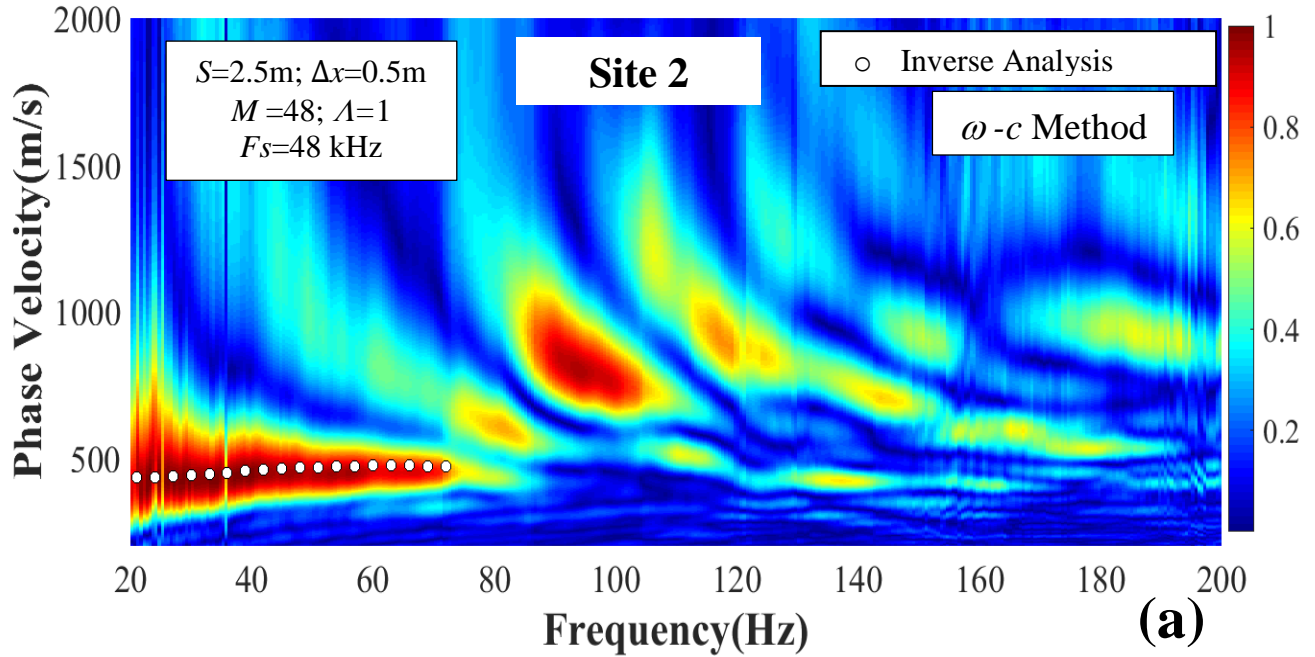

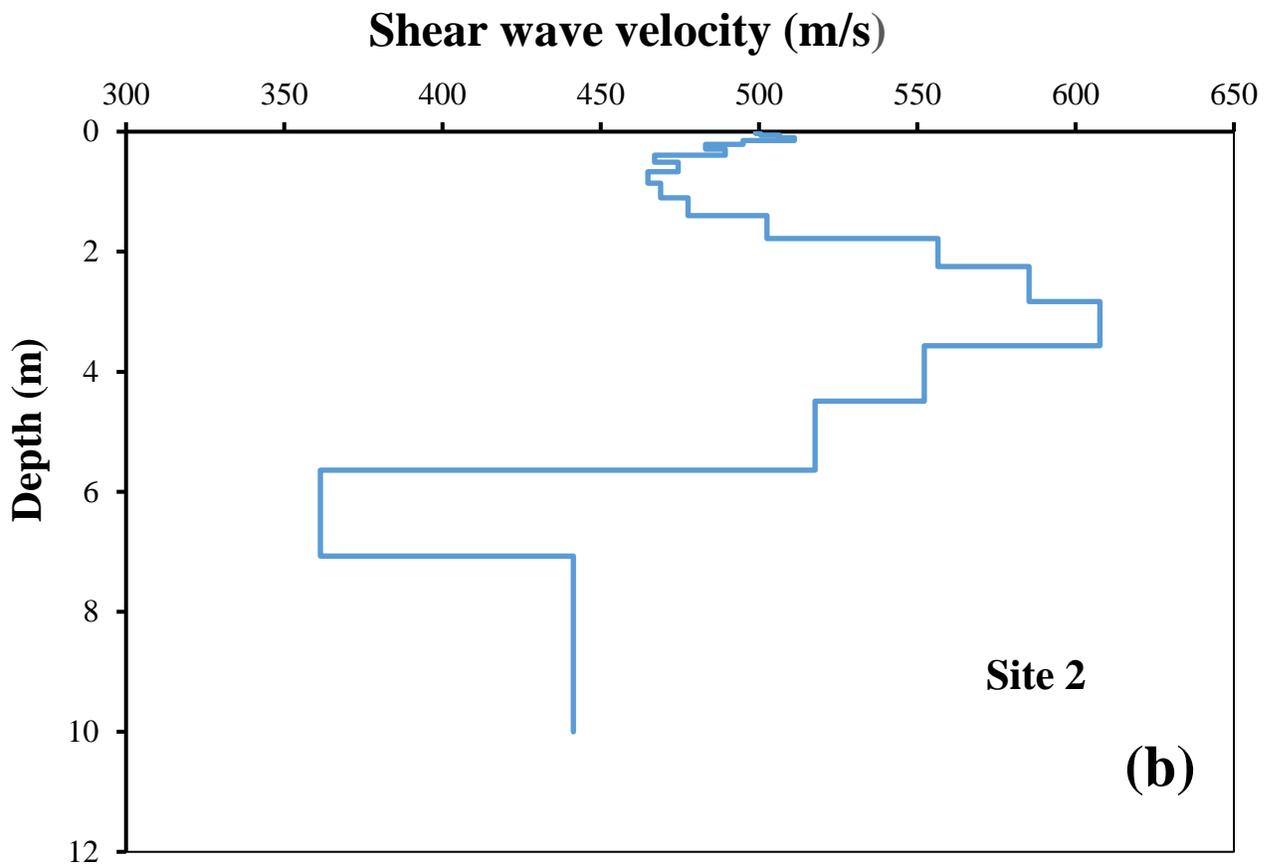

**Figure 19**. (a) A comparison between dispersion image and predicted theoretical curve from inverse analysis; (b) theoretical soil profile from inverse analysis for Site-2

**Table: 1.** Min. frequency (Hz) for fundamental mode determined qualitatively from different method and spread length using 24 receivers

| Method | Min. frequency (Hz) for different spread length ($X$) | | |
|---|---|---|---|
| | $X = 46$ m | $X = 92$ m | $X = 184$ m |
| $\omega$-$c$ | 15 | 13 | 9 |
| $\omega$-$\kappa$ | 15 | 13 | 10 |
| $\tau$-$p$ | 15 | 13 | 10 |

**Table: 2.** Min. frequency (Hz) for fundamental mode determined qualitatively from different number of receivers and spread length in $\omega$-$c$ method.

| Number of sensors ($M$) | Min. frequency (Hz) for different spread length ($X$) | | |
|---|---|---|---|
| | $X = 46$ m | $X = 92$ m | $X = 184$ m |
| 24 | 15 | 13 | 9 |
| 48 | 15 | 13 | 9 |
| 96 | 15 | 13 | 9 |